\documentclass[sigconf]{acmart}
\usepackage{booktabs} 
\usepackage{makecell}

\usepackage[T1]{fontenc}
\usepackage{epstopdf}
\usepackage{multirow}
\usepackage{balance}
\usepackage{amsfonts}
\usepackage{amsmath}
\usepackage{amssymb}
\usepackage{graphicx}
\usepackage{microtype}
\usepackage{url}
\usepackage{enumitem}
\usepackage{wrapfig,lipsum}
\usepackage{multirow}
\usepackage{makecell}
\usepackage{wrapfig}

\usepackage[boxed, ruled,vlined,linesnumbered]{algorithm2e}
\usepackage{bbold}
\usepackage{algpseudocode}
\usepackage{microtype}

\usepackage{threeparttable}
\usepackage{bm}
\usepackage{subfigure}
\usepackage{booktabs}
\usepackage{caption}

\newcommand{\bs}[1]{\boldsymbol{#1}}

\setcopyright{rightsretained}

\newtheorem{thm:def}{Definition}
\newtheorem{thm:eg}{Example}
\newtheorem{thm:lem}{Lemma}
\newtheorem{thm:obs}{Observation}
\newtheorem{thm:req}{Requirement}

\newcommand{\nop}[1]{}

\newcommand{\ie}{{\sl i.e.}}
\newcommand{\eg}{{\sl e.g.}}

\begin{document}

\copyrightyear{2018} 
\acmYear{2018} 
\setcopyright{acmcopyright}
\acmConference[CIKM '18]{The 27th ACM International Conference on Information and Knowledge Management}{October 22--26, 2018}{Torino, Italy}
\acmBooktitle{The 27th ACM International Conference on Information and Knowledge Management (CIKM '18), October 22--26, 2018, Torino, Italy}
\acmPrice{15.00}
\acmDOI{10.1145/3269206.3271737}
\acmISBN{978-1-4503-6014-2/18/10}

\fancyhead{}

\title{Weakly-Supervised Neural Text Classification}


\author{Yu Meng, Jiaming Shen, Chao Zhang, Jiawei Han}
\affiliation{%
  \institution{Department of Computer Science, University of at Illinois Urbana-Champaign, IL, USA}
}
\affiliation{%
  \institution{\{yumeng5, js2, czhang82, hanj\}@illinois.edu}
}

%
%
%

\renewcommand{\shorttitle}{Weakly-Supervised Neural Text Classification}
\renewcommand{\shortauthors}{Y. Meng et al.}


\begin{abstract}

Deep neural networks are gaining increasing popularity for the classic text
classification task, due to their strong expressive power and less requirement
for feature engineering.  Despite such attractiveness, neural text
classification models suffer from the lack of training data in many real-world
applications. Although many semi-supervised and weakly-supervised text
classification models exist, they cannot be easily applied to deep neural
models and meanwhile support limited supervision types.  In this paper, we
propose a weakly-supervised method that addresses the lack of training data in
neural text classification. Our method consists of two modules: (1) a
pseudo-document generator that leverages seed information to generate
pseudo-labeled documents for model pre-training, and (2) a self-training module
that bootstraps on real unlabeled data for model refinement.  Our method has
the flexibility to handle different types of weak supervision and can be easily
integrated into existing deep neural models for text classification.  We have
performed extensive experiments on three real-world datasets from different
domains. The results demonstrate that our proposed method achieves inspiring
performance without requiring excessive training data and outperforms baseline
methods significantly \footnote{Source code can be found at \url{https://github.com/yumeng5/WeSTClass}.}. 

\end{abstract}


\maketitle


\section{Introduction}

Text classification plays a fundamental role in a wide variety of applications,
ranging from sentiment analysis \cite{Tang2015DocumentMW} to document
categorization \cite{Yang2016HierarchicalAN} and query intent classification
\cite{Tsur2016IdentifyingWQ}.  Recently, deep neural models --- including
convolutional neural networks (CNNs) \cite{kim2014convolutional, zhang2015text,
  johnson2014effective, zhang2015character} and recurrent neural networks
(RNNs) \cite{socher2011semi, socher2011dynamic, Yang2016HierarchicalAN} ---
have demonstrated superiority for this classic task.  The attractiveness of
these neural models for text classification is mainly two-fold. First, they can
largely reduce feature engineering efforts by automatically learning
distributed representations that capture text semantics.  Second, they enjoy
strong expressive power to better learn from the data and yield better classification performance.

Despite the attractiveness and increasing popularity of neural models for text
classification, \emph{the lack of training data} is still a key bottleneck that
prohibits them from being adopted in many practical scenarios. Indeed, training
a deep neural model for text classification can easily consume million-scale
labeled documents. Collecting such training data requires domain experts to
read through millions of documents and carefully label them with domain
knowledge, which is often too expensive to realize.

To address the label scarcity bottleneck, we study the problem of learning
neural models for text classification \emph{under weak supervision}. In many
scenarios, while users cannot afford to label many documents for training
neural models, they can provide a small amount of seed information for the
classification task. Such seed information may arrive in various forms: either
a set of representative keywords for each class, or a few (less than a dozen)
labeled documents, or even only the surface names of the classes. Such a
problem is called \emph{weakly-supervised} text classification.

There have been many studies related to weakly-supervised text classification.
However, training neural models for text classification under weak supervision
remains an open research problem\nop{that is largely unsolved}.  Several
semi-supervised neural models have been proposed \cite{Miyato2016AdversarialTM,
  Xu2017VariationalAF}, but they still require hundreds or even thousands of
labeled training examples, which are not available in the weakly supervised
setting \cite{Oliver2018RealisticEO}. Along another line, there are existing
methods that perform weakly-supervised text classification, including latent
variable models \cite{li2016effective} and embedding-based methods
\cite{tang2015pte, li2018unsupervised}.  These models have the following
limitations: (1) \emph{supervision inflexibility}: they can only handle one
type of seed information, either a collection of labeled documents or a set of
class-related keywords, which restricts their applicabilities; (2) \emph{seed
  sensitivity}: the ``seed supervision'' from users completely controls the
model training process, making the learned model very sensitive to the initial
seed information; (3) \emph{limited extensibility}: these methods are specific
to either latent variable models or embedding methods, and cannot be readily
applied to learn deep neural models based on CNN or RNN.

\begin{figure*}[!t]
\includegraphics[width = 0.95\textwidth]{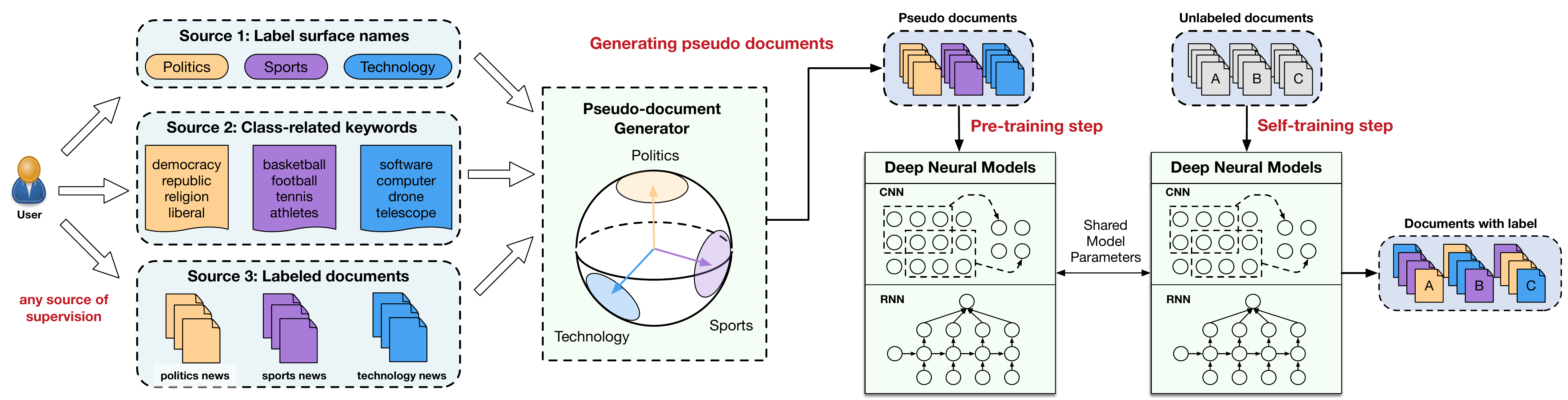}
\vspace{-2ex}
\caption{\textsc{WeSTClass}
  consists of two key modules: (1) a pseudo-document generator that leverages
  seed information to generate pseudo-labeled documents for model pre-training,
  and (2) a self-training module that bootstraps on real unlabeled data for
  model refinement.}
\label{fig:overview}
\vspace{-2ex}
\end{figure*}

We propose a new method, named \textsc{WeSTClass}, for \textbf{We}akly-\textbf{S}upervised 
\textbf{T}ext \textbf{Class}ification.  As shown in Figure \ref{fig:overview}, \textsc{WeSTClass}
contains two modules to address the above challenges.  The first module is a
pseudo-document generator, which leverages seed information to generate pseudo
documents as synthesized training data.  By assuming word and document
representations reside in the same semantic space, we generate pseudo documents
for each class by modeling the semantics of each class as a high-dimensional
spherical distribution \cite{fisher1953dispersion}, and further sampling
keywords to form pseudo documents. The pseudo document generator can not only
expand user-given seed information for better generalization, but also handle
different types of seed information (\eg, label surface names, class-related
keywords, or a few labeled documents) flexibly.



The second key module of our method is a self-training module
that fits real unlabeled documents for model refinement.
First, the self-training module uses pseudo documents to
pre-train either CNN-based or RNN-based models to produce an
initial model, which serves as a starting point in the subsequent
model refining process. Then, it applies a self-training
procedure, which iteratively makes predictions on real
unlabeled documents and leverages high-confidence predictions
to refine the neural model.

In summary, this paper makes the following contributions:
\begin{enumerate}[leftmargin=*] 
  \item We design \textsc{WeSTClass} method for addressing the label scarcity
    bottleneck of neural text classification.
    To the best of our knowledge, \textsc{WeSTClass} is the first
    weakly-supervised text classification method that can be applied to most
    existing neural models and meanwhile handle different types of seed
    information.

  \item We propose a novel pseudo document generator by modeling the class
    semantic as a spherical distribution. The generator is able to generate
    pseudo documents that are highly correlated to each class, and meanwhile
    effectively expands user-provided seed information for better
    generalization.

  \item We propose a self-training algorithm for training deep neural models by
    leveraging pseudo documents. The self-training algorithm can iteratively
    bootstrap the unlabeled data to obtain high-quality deep neural models, and
    is generic enough to be integrated into either CNN-based or RNN-based
    models.

\item We conduct a thorough evaluation of our proposed method on three
  real-world datasets from different domains. The experiment results show that
  our method can achieve inspiring text classification performance even without
  excessive training data and outperforms various baselines. 

\end{enumerate}



\section{Related Work}

In this section, we review existing studies for weakly-supervised text classification, which can be categorized into two classes: (1)
latent variable models; and (2) embedding-based models.

\subsection{Latent Variable Models}

Existing latent variable models for weakly-supervised text classification mainly extend topic models by incorporating user-provided seed information.
Specifically, semi-supervised PLSA \cite{lu2008opinion} extends the classic PLSA model by incorporating a conjugate prior based on expert review segments (topic keywords or phrases) to force extracted topics to be aligned with provided review segments.  
\cite{ganchev2010posterior} encodes prior knowledge and indirect supervision in constraints on posteriors of latent variable probabilistic models. 
Descriptive LDA \cite{Chen2015DatalessTC} uses an LDA
model as the describing device to infer Dirichlet priors from given category
labels and descriptions. The Dirichlet priors guides LDA to induce the
category-aware topics.  Seed-guided topic model \cite{li2016effective} takes a
small set of seed words that are relevant to the semantic meaning of the
category, and then predicts the category labels of the documents through two
kinds of topic influence: category-topics and general-topics. The labels of the
documents are inferred based on posterior category-topic assignment. 
Our method
differs from these latent variable models in that it is a weakly-supervised
neural model. As such, it enjoys two advantages over these latent variable
models: (1) it has more flexibility to handle different types of seed
information which can be a collection of labeled documents or a set of seed keywords
related to each class; (2) it does not need to impose assumptions on
document-topic or topic-keyword distributions, but instead directly uses massive data
to learn distributed representations to capture text semantics.


\subsection{Embedding-based Models}

Embedding-based weakly supervised models use seed information to derive
vectorized representations for documents and label names for the text classification
task. Dataless classification \cite{chang2008importance, song2014dataless}
takes category names and projects each word and document into the same semantic
space of Wikipedia concepts. Each category is represented with words in the
category label. The document classification is performed based on the vector
similarity between a document and a category using explicit semantic analysis \cite{Gabrilovich2007ComputingSR}.
Unsupervised neural categorization \cite{li2018unsupervised} takes category
names as input and applies a cascade embedding approach: First the seeded
category names and other significant phrases (concepts) are embedded into
vectors for capturing concept semantics. Then the concepts are embedded into a
hidden category space to make the category information explicit. Predictive
text embedding \cite{tang2015pte} is a semi-supervised algorithm that utilizes
both labeled and unlabeled documents to learn text embedding specifically for a
task. Labeled data and different levels of word co-occurrence information are
first represented as a large-scale heterogeneous text network and then embedded
into a low dimensional space that preserves the semantic similarity of words
and documents. Classification is performed by using one-vs-rest logistic
regression model as classifier and the learned embedding as input.  Compared
with our method, these embedding-based weakly supervised methods cannot be
directly applied to deep neural models (CNN, RNN) for the text classification
task. Furthermore, while they allow the seed information to directly control
the model training process, we introduce a pseudo document generation paradigm which is generalized from the seed information.
Hence, our model is less prone to seed information overfitting and enjoys
better generalization ability.




\section{Preliminaries}

In this section, we formulate the problem of weakly-supervised text
classification, and give an overview of our proposed method.

\subsection{Problem Formulation}

Given a text collection $\mathcal{D} = \{D_{1}, \dots, D_{n} \}$ and $m$ target
classes $\mathcal{C} = \{C_{1}, \dots, C_{m}\}$, text classification aims to
assign a class label $C_{j} \in \mathcal{C}$ to each document $D_{i} \in
\mathcal{D}$.  To characterize each class, traditional supervised text
classification methods rely on large amounts of labeled documents.  In this
work, we focus on the text classification under weakly-supervised setting where
the supervision signal comes from one of the following sources: (1) \emph{label
  surface names}: $\mathcal{L} = \{L_{j}\}|_{j=1}^{m}$, where $L_{j}$ is the
surface name for class $C_{j}$, (2) \emph{class-related keywords}: $\mathcal{S}
= \{S_{j}\}|_{j=1}^{m}$, where $S_{j} = \{w_{j,1}, \dots, w_{j,k}\}$ represents
a set of $k$ keywords in class $C_{j}$, and (3) \emph{labeled documents}:
$\mathcal{D}^{L} = \{ \mathcal{D}_{j}^{L} \}|_{j=1}^{m}$, where
$\mathcal{D}_{j}^{L} = \{D_{j,1}, \dots, D_{j,l} \}$ denotes a set of $l$ ($l
\ll n$) labeled documents in class $C_{j}$.  In many scenarios, the above weak
supervision signals can be easily obtained from users. Finally, we define our
problem as follows:

\begin{definition} [Problem Formulation] Given a text collection $\mathcal{D} =
  \{D_{1}, \dots, D_{n} \}$, target classes $\mathcal{C} = \{C_{1}, \dots,
  C_{m}\}$, and weak supervision from either $\mathcal{L}$, $\mathcal{S}$ or
  $\mathcal{D}^{L}$, the weakly-supervised text classification task aims to
  assign a label $C_{j} \in \mathcal{C}$ to each $D_{i} \in \mathcal{D}$.
\end{definition}

\subsection{Method Overview}

Our proposed weakly-supervised text classification method
contains two key modules.  The first one is a pseudo-document
generator that unifies seed information and outputs pseudo
documents for model training.  We assume words and documents
share a joint semantic space which provides flexibility for
handling different types of seed information.  Then, we model
each class as a high-dimensional spherical distribution from
which keywords are sampled to form pseudo documents as training
data.  The second key module of our method is a self-training
module that can be easily integrated into existing deep neural
models, either CNN-based or RNN-based.  It first uses the
generated pseudo documents to pre-train neural models, which
allows the model to start with a good initialization.  Then, a
self-training procedure is applied to iteratively refine the
neural model using unlabeled real documents  based on the
model's high-confidence predictions.  We show the entire
process of our method in Figure \ref{fig:overview}.

%


\section{Pseudo Document Generation}
\label{gen}


In this section, we describe the details of the pseudo-document generator,
which leverages seed information to generate a bunch of pseudo documents that
are correlated to each class. Below, we first introduce how to model class distributions in a
joint semantic space with words and documents, and then describe the pseudo document generation process.


\subsection{Modeling Class Distribution}

To effectively leverage user-provided seed information and capture the semantic correlations between words, documents and classes, we assume words and documents share a joint semantic
space, based on which we learn a generative model for each class to generate
pseudo documents.

Specifically, we first use the Skip-Gram model \cite{mikolov2013distributed} to learn
$p$-dimensional vector representations of all the words in the corpus.
Furthermore, since directional similarities between vectors are more effective
in capturing semantic correlations \cite{sra2016directional,
  banerjee2005clustering, levy2015improving}, we normalize all the
$p$-dimensional word embeddings so that they reside on a unit sphere in $\mathbb{R}^{p}$, which is the joint semantic space. We call it ``joint'' because we assume pseudo document vectors reside on the same unit sphere as well, which we will explain in Section \ref{sec:gen}.
 We retrieve a set of keywords in the semantic space that are correlated to each class
based on the seed information.  We describe how to handle different types of
seed information as follows:

\begin{itemize}[leftmargin=0.5cm] 

  \item \textbf{Label surface names}: When only label surface names $\mathcal{L}$ are given as seed
    information, for each class $j$ we use the embedding of its surface name $L_{j}$ to
    retrieve top-$t$ nearest words in the semantic space. We set $t$ to be the largest number that does not results in shared words across different classes.     


  \item \textbf{Class-related keywords}: When users provide a list of related
    keywords $S_j$ for each class $j$, we use the
    embeddings of these seed keywords to find top-$t$ keywords in the semantic
    space, by measuring the average similarity to the seed keywords.


  \item \textbf{Labeled documents}: When users provide a small number of
    documents $\mathcal{D}^{L}_{j}$ that are correlated with class $j$, we first extract $t$ representative keywords in $\mathcal{D}^{L}_{j}$ using tf-idf weighting, and then consider them as class-related keywords.


\end{itemize}





After obtaining a set of keywords that are correlated with each class, we model
the semantic of each class as a von Mises Fisher (vMF) distribution
\cite{banerjee2005clustering, gopal2014mises}, which models word embeddings on
a unit sphere in $\mathbb{R}^{p}$ and has been shown effective for various
tasks \cite{BatmanghelichSN16, ZhangLLYZH017}.  Specifically, we define the probability distribution of a class as: 
$$
f(\bs{x};\bs{\mu},\kappa) = c_p(\kappa)e^{\kappa\bs{\mu}^T\bs{x}},
$$
where $\kappa \ge 0$, $\|\bs{\mu}\| = 1$, $p \ge 2$ and the normalization constant $c_p(\kappa)$ is given by 
$$
c_p(\kappa) = \frac{\kappa^{p/2-1}}{(2\pi)^{p/2} I_{p/2-1}(\kappa)},
$$
where $I_r(\cdot)$ represents the modified Bessel function of the first kind at
order $r$.  We justify our choice of the vMF distribution as follows: the vMF
distribution has two parameters---the mean direction $\bs{\mu}$ and the
concentration parameter $\kappa$. The distribution of keywords on the unit
sphere for a specific class concentrates around the mean direction $\bs{\mu}$, and is
more concentrated if $\kappa$ is large.   Intuitively, the mean direction
$\bs{\mu}$ acts as a semantic focus on the unit sphere, and produces relevant
semantic embeddings around it, where concentration degree is controlled by the
parameter $\kappa$.

Now that we have leveraged the seed information to obtain a set of keywords for
each class on the unit sphere, we can use these correlated keywords to fit a
vMF distribution $f(\bs{x};\bs{\mu},\kappa)$. Specifically, let $X$ be a set of
vectors for the keywords on the unit sphere, i.e., 
$$
X = \{\bs{x}_i \in \mathbb{R}^{p} \mid \bs{x}_i \text{ drawn from }  f(\bs{x};\bs{\mu},\kappa), 1\le i \le t\},
$$
then we use the maximum likelihood estimates \cite{sra2012short,
  banerjee2005clustering} for finding the parameters $\bs{\hat{\mu}}$ and
$\hat{\kappa}$ of the vMF distribution: 
$$
\bs{\hat{\mu}} = \frac{\sum_{i=1}^{t} \bs{x}_i}{\|\sum_{i=1}^{t} \bs{x}_i\|},
$$
and
$$
\frac{I_{p/2}(\hat{\kappa})}{I_{p/2-1}(\hat{\kappa})} = \frac{\|\sum_{i=1}^{t} \bs{x}_i\|}{t}.
$$
Obtaining an analytic solution for $\hat{\kappa}$ is infeasible because the
formula involves an implicit equation which is a ratio of Bessel functions. We
thus use a numerical procedure  based on Newton's method
\cite{banerjee2005clustering} to derive an approximation of $\hat{\kappa}$.

\subsection{Generating Pseudo Documents}
\label{sec:gen}
To generate a pseudo document $D_i^*$ (we use $D_i^*$ instead of $D_i$ to
denote it is a pseudo document) of class $j$, we propose a generative mixture
model based on class $j$'s distribution $f(\bs{x};\bs{\mu}_j,\kappa_j)$.  The
mixture model repeatedly generates a number of terms to form a pseudo document;
when generating each term, the model chooses from a background distribution
with probability $\alpha$ ($0 < \alpha < 1$) and from the class-specific
distribution with probability $1-\alpha$. 


The class-specific distribution is defined based on class
$j$'s distribution $f(\bs{x};\bs{\mu}_j,\kappa_j)$. Particularly, we first
sample a document vector $\bs{d}_i$ from $f(\bs{x};\bs{\mu}_j,\kappa_j)$, then
build a keyword vocabulary $V_{d_i}$ for $\bs{d}_i$ that contains the top-$\gamma$
words with most similar word embedding with $\bs{d}_i$. These $\gamma$ words in
$V_{d_i}$ are highly semantically relevant with the topic of pseudo document
$D_i^*$ and will appear frequently in $D_i^*$. Each term of a
pseudo document is generated according to the following probability distribution:

\begin{equation} \label{eq:1}
p(w \mid \bs{d}_i) = \begin{cases}
\alpha p_B(w) & w \notin V_{d_i} \\
\alpha p_B(w) + (1-\alpha) \frac{\exp(\bs{d}_i^T \bs{v}_w)}{\sum_{w' \in V_{d_i}} \exp(\bs{d}_i^T \bs{v}_{w'})} & w \in V_{d_i}
\end{cases}
\end{equation}
where $\bs{v}_w$ is the word embedding for $w$ and $p_B(w)$ is the background
distribution for the entire corpus.

Note that we generate document vectors from $f(\bs{x};\bs{\mu}_j,\kappa_j)$ instead of fixing them to be $\bs{\mu}_j$.
The reason is that some class (\eg, Sports) may cover a wide range of topics (\eg, athlete activities, sport competitions, etc.), but using $\bs{\mu}_j$ as the pseudo document vector will only attract words that are semantically similar to the centroid direction of a class. 
Sampling pseudo document vectors from the distribution, however, allows the generated pseudo documents to be more semantically diversified and thus cover more information about the class. 
Consequently, models trained on such more diversified pseudo documents are expected to have better generalization ability. 


Algorithm \ref{alg:pseudoDocumentsGeneration} shows the whole process of generating a collection of $\beta$ pseudo documents per class. For each class $j$, given the learned class distributions and the average length of pseudo documents $dl$\footnote{The length of each pseudo document can be either manually set or equal to the average document length in the real document collection.}, we draw a document vector $\bs{d}_{i}$ from class $j$'s distribution $f(\bs{x};\bs{\mu}_j,\kappa_j)$. After that, we generate $dl$ words sequentially based on $\bs{d}_{i}$ and add the generated document into the pseudo document collection $\mathcal{D}^{*}_j$ of class $j$. After the above process repeats $\beta$ times, we finally obtain $\mathcal{D}^{*}_j$ which contains $\beta$ pseudo documents for class $j$. 
  
  \begin{algorithm}[!t]
  \caption{Pseudo Documents Generation.}
  \label{alg:pseudoDocumentsGeneration}
  \KwIn{
  Class distributions $\{ f(\bs{x};\bs{\mu}_j,\kappa_j) \} | _{j=1}^{m}$; average document length $dl$; number of pseudo documents $\beta$ to generate for each class.
  }
  \KwOut{A set of $m \times \beta$ pseudo documents $\mathcal{D^{*}}$.}
  Initialize $\mathcal{D^{*}} \gets \emptyset$\;
  \For{class index j from 1 to m} {
  	Initialize $\mathcal{D}^{*}_j \gets \emptyset$\;
	  \For{pseudo document index i from 1 to $\beta$} {
		Sample document vector $\bs{d}_i$ from $f(\bs{x};\bs{\mu}_j,\kappa_j)$\;
	  	$D_{i}^{*} \gets $ empty string\;
		\For{word index $k$ from $1$ to $dl$} {
			Sample word $w_{i,k} \sim p(w \mid \bs{d}_i)$ based on Eq. (\ref{eq:1})\; 
			$D_{i}^{*} = D_{i}^{*} \oplus w_{i,k}$ // concatenate $w_{i,k}$ after $D_{i}^{*}$\;
		}
		$\mathcal{D}^{*}.append(D_{i}^{*})$\;
	  }
  	$\mathcal{D}^{*} \gets \mathcal{D}^{*} \cup \mathcal{D}^{*}_j$\;
	}
  Return $\mathcal{D}^{*}$\;
\end{algorithm}











\section{Neural Models with Self-Training}
\label{sec: self-train}

In this section, we present the self-training module that trains deep neural
models with the generated pseudo documents. The self-training module  first uses
the pseudo documents to pre-train a deep neural network, and then iteratively
refines the trained model on the real unlabeled documents in a bootstrapping
fashion. In the following, we first present the pre-training and the
self-training steps in Section \ref{sect:pretrain} and \ref{sect:self-train}, and then
demonstrate how the framework can be instantiated with CNN and RNN models in
Section \ref{sect:instant}.


\subsection{Neural Model Pre-training}
\label{sect:pretrain}



As we have obtained pseudo documents for each class, we use them to pre-train a
neural network $M$\footnote{When the supervision source is \textbf{labeled documents}, these seed documents will be used to augment the pseudo document
  set during the pre-training step.}.  A naive way of creating the label for a
pseudo document $D_i^*$ is to directly use the associated class label that
$D_i^*$ is generated from, \emph{i.e.} using one-hot encoding where the
generating class takes value $1$ and all other classes are set to $0$. However,
this naive strategy often causes the neural model to overfit to the pseudo
documents and have limited performance when classifying real documents, due to the fact that the generated pseudo documents do not contain word ordering information. To
tackle this problem, we create pseudo labels for pseudo documents. 
In Equation (\ref{eq:1}), we design pseudo documents to be generated from a mixture of
background and class-specific word distributions, controlled by a balancing
parameter $\alpha$. Such a process naturally leads to our design of the
following procedure for pseudo label creation: we evenly split the fraction of
the background distribution into all $m$ classes, and set the pseudo label
$\bs{l}_i$ for pseudo document $D_i^*$ as
$$
l_{ij} = \begin{cases}
(1-\alpha) + \alpha/m & \text{$D_i^*$ is generated from class $j$} \\
\alpha/m & \text{otherwise}
\end{cases}
$$

After creating the pseudo labels, we pre-train a neural model $M$ by generating
$\beta$ pseudo documents for each class, and minimizing the KL divergence loss from
the neural network outputs $Y$ to the pseudo labels $L$,
namely
$$
loss = KL(L\|Y) = \sum_i \sum_j l_{ij} \log \frac{l_{ij}}{y_{ij}}
$$
We will detail how we instantiate the neural model $M$ shortly in Section
\ref{sect:instant}.

\subsection{Neural Model Self-training}
\label{sect:self-train}

While the pre-training step produces an initial neural model $M$, the
performance of the $M$ is not the best one can hope for. The major reason is
that the pre-trained model $M$ only uses the set of pseudo documents but fails to
take advantage of the information encoded in the real unlabeled documents.  
The self-training step is designed to tackle the above issues.  Self-training
\cite{nigam2000analyzing,rosenberg2005semi} is a common strategy used in
classic semi-supervised learning scenarios. The rationale behind self-training
is to first train the model with labeled data, and then bootstrap the learning
model with its current highly-confident predictions. 

After the pre-training step, we use the pre-trained model to classify all
unlabeled documents in the corpus and then apply a self-training strategy to
improve the current predictions. During self-training, we iteratively compute
pseudo labels based on current predictions and refine model parameters by
training the neural network with pseudo labels. Given the current outputs $Y$,
the pseudo labels are computed using the same self-training formula as in
\cite{xie2016unsupervised}:
$$
l_{ij} = \frac{y_{ij}^2/f_{j}}{\sum_{j'}y_{ij'}^2/f_{j'}}
$$
where $f_{j} = \sum_{i}y_{ij}$ is the soft frequency for class $j$.

Self-training is performed by iteratively computing pseudo labels and
minimizing the KL divergence loss from the current predictions $Y$ to the
pseudo labels $L$. This process terminates when less than $\delta\%$ of the
documents in the corpus have class assignment changes.

Although both pre-training and self-training create pseudo labels and use them
to train neural models, it is worth mentioning the difference between them: in
pre-training, pseudo labels are paired with generated pseudo documents to
distinguish them from given labeled documents (if provided) and prevent the
neural models from overfitting to pseudo documents; in self-training,
pseudo labels are paired with every unlabeled real documents from corpus and
reflect current high confidence predictions.

\subsection{Instantiating with CNNs and RNNs}
\label{sect:instant}

As mentioned earlier, our method for text classification is generic enough to
be applied to most existing deep neural models. In this section, we instantiate
the framework with two mainstream deep neural network models: convolution
neural networks (CNN) and recurrent neural networks (RNN), by focusing on how
they are used to learn document representations and perform classification.

\subsubsection{CNN-Based Models}

CNNs have been explored for text classification \cite{kim2014convolutional}.
When instantiating our framework with CNN, the input to a CNN is a document
of length $dl$ represented by a concatenation of word vectors,
i.e.,
$$
\bs{d} = \bs{x}_1 \oplus \bs{x}_2 \oplus \dots \oplus \bs{x}_{dl},
$$ 
where $\bs{x}_i \in \mathbb{R}^{p}$ is
the $p$ dimensional word vector of the $i$th word in the document. We use
$\bs{x}_{i:i+j}$ to represent the concatenation of word vectors $\bs{x}_i,
\bs{x}_{i+1}, \dots, \bs{x}_{i+j}$. For window size of $h$, a feature $c_i$ is
generated from a window of words $\bs{x}_{i:i+h-1}$ by the following
convolution operation
$$
c_i = f(\bs{w} \cdot \bs{x}_{i:i+h-1} + b),
$$ where $b \in \mathbb{R}$ is a bias term, $\bs{w} \in \mathbb{R}^{hp}$ is the
filter operating on $h$ words.
For each possible size-$h$ window of words, a feature map is generated as
$$
\bs{c} = [c_1, c_2, \dots, c_{dl-h+1}].
$$
Then a max-over-time pooling operation is performed on $\bs{c}$ to output the
maximum value $\hat{c} = \max(\bs{c})$ as the feature corresponding to this
particular filter. If we use multiple filters, we will obtain multiple features
that are passed through a fully connected softmax layer whose output is the
probability distribution over labels.

\subsubsection{RNN-Based Models}

Besides CNNs, we also discuss how to instantiate our framework with RNNs. We
choose the Hierarchical Attention Network (HAN) \cite{yang2016hierarchical} as
an exemplar RNN-based model. HAN consists of sequence encoders and attention layers for both
words and sentences. In our context, the input document is represented by a
sequence of sentences $s_i, i\in[1,L]$ and each sentence is represented by a
sequence of words $w_{it}, t\in[1,T]$. At time $t$, the GRU
\cite{Bahdanau2014NeuralMT} computes the new state as
$$
\bs{h}_t = (\bs{1} - \bs{z}_t) \odot \bs{h}_{t-1} + \bs{z}_t \odot \tilde{\bs{h}}_t,
$$
where the update gate vector
$$
\bs{z}_t = \sigma (W_z \bs{x}_t + U_z \bs{h}_{t-1} + \bs{b}_z),
$$
the candidate state vector
$$
\tilde{\bs{h}}_t = \tanh (W_h \bs{x}_t + \bs{r}_t \odot (U_h \bs{h}_{t-1}) + \bs{b}_h),
$$
the reset gate vector
$$
\bs{r}_t = \sigma (W_r \bs{x}_t + U_r \bs{h}_{t-1} + \bs{b}_r),
$$
and $\bs{x}_t$ is the sequence vector (word embedding or sentence vector) at
time $t$.  After encoding words and sentences, we also impose the attention
layers to extract important words and sentences with the attention mechanism,
and derive their weighted average as document representations.




\section{Experiments}
\label{sec:exp}

In this section, we evaluate the empirical performance of our method
for weakly supervised text classification.

\subsection{Datasets}

We use three corpora from different domains to evaluate the performance of our
proposed method: (1) \textbf{The New York Times}: We crawl $13,081$ news
articles using the New York Times
API\footnote{\url{http://developer.nytimes.com/}}. This corpus covers $5$ major
news topics; (2) \textbf{AG's News}: We use the same AG's News dataset from
\cite{zhang2015character} and take its training set portion ($120,000$
documents evenly distributed into $4$ classes) as the corpus for evaluation;
(3) \textbf{Yelp Review}: We use the Yelp reviews polarity dataset from
\cite{zhang2015character} and take its testing set portion ($38,000$ documents
evenly distributed into $2$ classes) as the corpus for evaluation.
Table~\ref{tab:dataset} provides the details of these datasets.

\begin{table*}[!t]
\caption{Dataset Statistics.}
\vspace{-0.3cm}
\label{tab:dataset}
\scalebox{0.95}{
\begin{tabular}{cccc}
\toprule
Corpus name & Classification type & Class name (Number of documents in the class) & Average document length \\
\midrule
The New York Times & Topic & Politics ($1451$), Arts ($1043$), Business ($1429$), Science ($519$), Sports ($8639$) & $778$ \\
AG's News & Topic & Politics ($30000$), Sports ($30000$), Business ($30000$), Technology ($30000$) & $45$ \\
Yelp Review & Sentiment & Good ($19000$), Bad ($19000$) & $155$ \\
\bottomrule
\end{tabular}
}
\end{table*}

\subsection{Baselines}

We compare \textsc{WeSTClass} with a wide range of baseline models, described as
follows.


\begin{itemize}[leftmargin=0.5cm] 
\item \textbf{IR with tf-idf}: this method accepts either \textbf{label surface name} or \textbf{class-related keywords} as  supervision. We treat the label name or keyword set for each class as a query, and score the relevance of document to this class using the tf-idf model. The class with highest relevance score is assigned to the document. 

\item \textbf{Topic Model}: this method accepts either \textbf{label surface
    name} or \textbf{class-related keywords} as supervision. We first train the
  LDA model \cite{blei2003latent} on the entire corpus. Given a document, we
  compute the likelihood of observing label surface names or the average
  likelihood of observing class-related keywords. The class with maximum
  likelihood will be assigned to the document. 

\item \textbf{Dataless \cite{chang2008importance,song2014dataless}}: this
  method 
\footnote{\url{https://cogcomp.org/page/software_view/Descartes}}
  accepts only \textbf{label surface name} as supervision. It leverages
  Wikipedia and uses Explicit Semantic Analysis
  \cite{Gabrilovich2007ComputingSR} to derive vector representations of both
  labels and documents.  The final document class is assigned based on the
  vector similarity between labels and documents. 

\item \textbf{UNEC \cite{li2018unsupervised}}: this method takes \textbf{label
    surface name} as its weak supervision. It categorizes documents by learning
  the semantics and category attribution of concepts inside the corpus. 
We use the authors' original implementation of this model. 

\item \textbf{PTE \cite{tang2015pte}}: this method
\footnote{\url{https://github.com/mnqu/PTE}}
  uses \textbf{labeled documents} as supervision. It first utilizes both labeled and unlabeled data to learn text embedding and then applies logistic regression model as classifier for text classification.   

\item \textbf{CNN \cite{kim2014convolutional}}: the original CNN model is a supervised text classification model and we extend it to incorporate all three types of supervision sources. If \textbf{labeled documents} are given, we directly train CNN model on the given labeled documents and then apply it on all unlabeled documents. If \textbf{label surface names} or \textbf{class-related keywords} are given, we first use the above ``IR with tf-idf'' or ``Topic Modeling'' method (depending on which one works better) to label all unlabeled documents. 
Then, we select $\beta$ labeled documents per class to pre-train CNN. Finally, we apply the same self-training module as described in Section \ref{sec: self-train} to obtain the final classifier. 

\item \textbf{HAN \cite{yang2016hierarchical}}: similar to the above CNN model,
  we extend the original HAN model
  \footnote{\url{https://github.com/richliao/textClassifier}} to incorporate
  all three types of supervision sources. 

\item \textbf{NoST-(CNN/HAN)}: this is a variant of \textsc{WeSTClass} without
  the self-training module, \ie, after pre-training CNN or HAN with pseudo
  documents, we directly apply it to classify unlabeled documents. 

\item \textbf{\textsc{WeSTClass}-(CNN/HAN)}: this is the full version of our proposed framework, with both pseudo-document generator and self-training module enabled. 

\end{itemize}

\subsection{Experiment Settings}
\label{sec:exp_setting}



We first describe our parameter settings as follows.  For all datasets, we use
the Skip-Gram model \cite{mikolov2013distributed} to train $100$-dimensional word
embeddings on the corresponding corpus. We set the background word distribution
weight $\alpha = 0.2$, the number of pseudo documents per class for pre-training $\beta =
500$, the size of class-specific vocabulary $\gamma = 50$ and the self-training
stopping criterion $\delta = 0.1$.

We apply our proposed framework on two types of state-of-the-art text
classification neural models: (1) CNN model, whose filter window sizes are $2,
3, 4, 5$ with $20$ feature maps each. (2) HAN model, which uses a forward GRU
with $100$ dimension output for both word and sentence encoding. Both the
pre-training and the self-training steps are performed using SGD with batch
size $256$.

The seed information we use as weak supervision for different datasets are
described as follows: (1) When the supervision source is \textbf{label surface
  name}, we directly use the label surface names of all classes; (2) When the
supervision source is \textbf{class-related keywords}, we manually choose $3$
keywords which do not include the class label name for each class.  The
selected keywords are shown in Tables \ref{tab:nyt_key}, \ref{tab:agnews_key}
and \ref{tab:yelp_key}, and we evaluate how our model is sensitive to such seed
keyword selection in Section \ref{subsec:keyword_sensitivity}; (3) When the
supervision source is \textbf{labeled documents}, we randomly sample $c$
documents of each class from the corpus ($c = 10$ for \textbf{The New York
  Times} and \textbf{AG's News}; $c = 20$ for \textbf{Yelp Review}) and use
them as the given labeled documents.  To alleviate the randomness, we repeat
the document selection process 10 times and show the performances with average
and standard deviation values.


\begin{table}
\caption{Keyword Lists for The New York Times Dataset.}
\vspace{-0.3cm}
\label{tab:nyt_key}
\scalebox{0.9}{
\begin{tabular}{cc}
\toprule
Class & Keyword List \\
\midrule
Politics & \{democracy, religion, liberal\}\\
Arts & \{music, movie, dance\} \\
Business & \{investment, economy, industry\} \\
Science & \{scientists, biological, computing\} \\
Sports & \{hockey, tennis, basketball\} \\
\bottomrule
\end{tabular}
}
\vspace{-0.3cm}
\end{table}

\begin{table}
\caption{Keyword Lists for AG's News Dataset.}
\vspace{-0.3cm}
\label{tab:agnews_key}
\scalebox{0.9}{
\begin{tabular}{cc}
\toprule
Class & Keyword List \\
\midrule
Politics & \{government, military, war\} \\
Sports & \{basketball, football, athletes\} \\
Business & \{stocks, markets, industries\} \\
Technology & \{computer, telescope, software\} \\
\bottomrule
\end{tabular}
}
\vspace{-0.3cm}
\end{table}

\begin{table}
\caption{Keyword Lists for Yelp Review Dataset.}
\vspace{-0.3cm}
\label{tab:yelp_key}
\scalebox{0.9}{
\begin{tabular}{cc}
\toprule
Class & Keyword List \\
\midrule
Good & \{terrific, great, awesome\} \\
Bad & \{horrible, disappointing, subpar\} \\
\bottomrule
\end{tabular}
}
\vspace{-0.3cm}
\end{table}

\subsection{Experiment Results}

In this subsection, we report our experimental results and our findings.

\subsubsection{Overall Text Classification Performance} In the first set of
experiments, we compare the classification performance of our method against all
the baseline methods on the three datasets. Both macro-F1 and micro-F1 metrics
are used to quantify the performance of different methods.  As shown in Tables
\ref{tab:macro} and \ref{tab:micro}, our proposed framework achieves the
overall best performances among all the baselines on three datasets with
different weak supervision sources. Specifically, in almost every case, \textbf{\textsc{WeSTClass}-CNN} yields the best performance among all methods; \textbf{\textsc{WeSTClass}-HAN} performs slightly worse than
\textbf{\textsc{WeSTClass}-CNN} but still outperforms other baselines.
We discuss the effectiveness of \textbf{\textsc{WeSTClass}} from the following aspects:

\begin{enumerate}[leftmargin=*] 
	
        \item When \textbf{labeled documents} are given as the supervision
          source, the standard deviation values of \textbf{\textsc{WeSTClass}-CNN} and
          \textbf{\textsc{WeSTClass}-HAN} are smaller than those of \textbf{CNN} and
          \textbf{HAN}, respectively. This shows that \textbf{\textsc{WeSTClass}} can
          effectively reduce the seed sensitivity and improve the robustness of
          CNN and HAN models.
	
        \item When the supervision source is \textbf{label surface name} or
          \textbf{class-related keywords}, we can see that \textbf{\textsc{WeSTClass}-CNN}
          and \textbf{\textsc{WeSTClass}-HAN} outperform \textbf{CNN} and \textbf{HAN},
          respectively. This demonstrates that pre-training with generated
          pseudo documents results in a better neural model initialization
          compared to pre-training with documents that are labeled using either
          \textbf{IR with tf-idf} or \textbf{Topic Modeling}.
	
        \item \textbf{\textsc{WeSTClass}-CNN} and \textbf{\textsc{WeSTClass}-HAN} always outperform
          \textbf{NoST-CNN} and \textbf{NoST-HAN}, respectively. Note that the
          only difference between \textbf{\textsc{WeSTClass}-CNN}/\textbf{\textsc{WeSTClass}-HAN} and
          \textbf{NoST-CNN}/\textbf{NoST-HAN} is that the latter two do not
          include the self-training module. The performance gaps between them
          thus clearly demonstrate the effectiveness of our self-training
          module. 
	
\end{enumerate}

\subsubsection{Effect of self-training module} 
\begin{figure}[t]
\vspace{-0.3cm}
\subfigure[\textsc{WeSTClass}-CNN -- New York Times]{
\includegraphics[width = 0.225\textwidth]{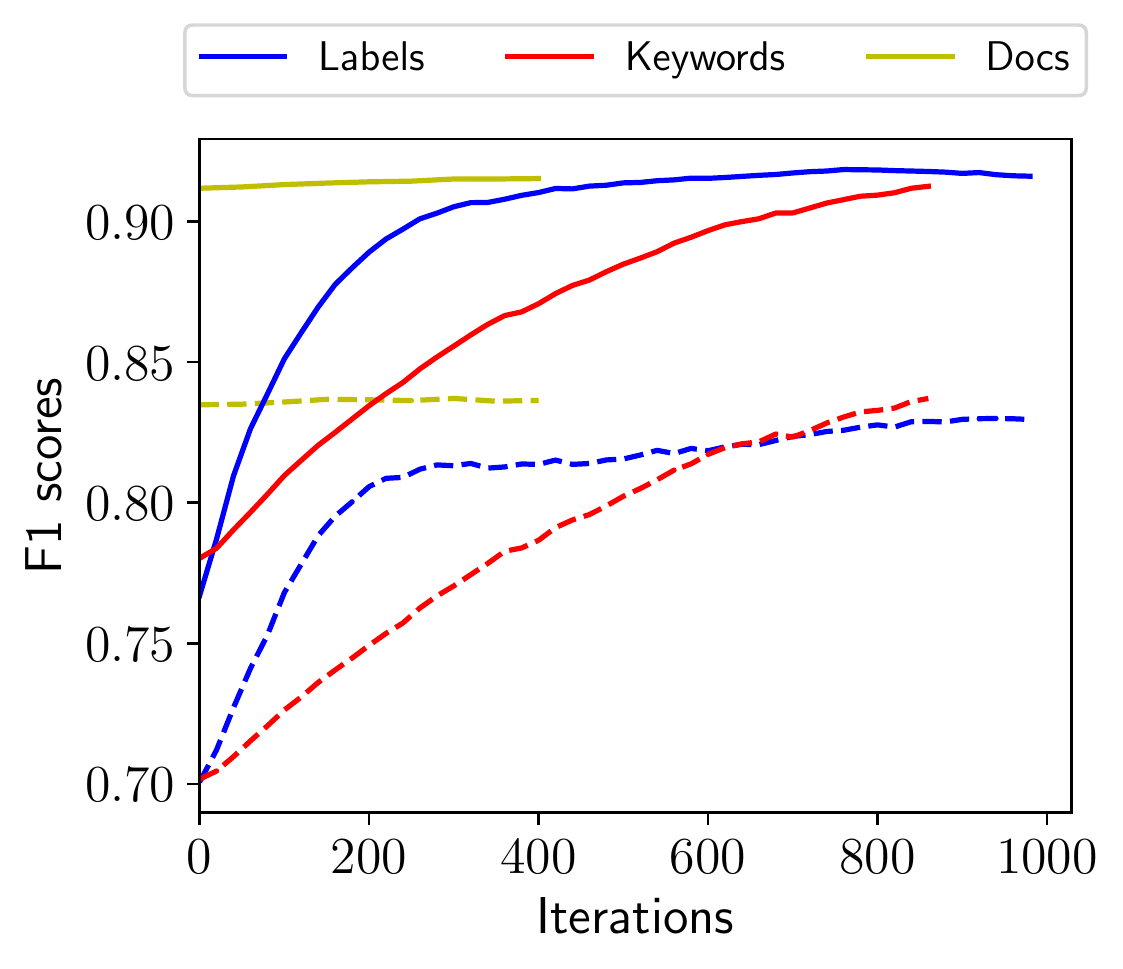}
}
\vspace{-0.1cm}
\subfigure[\textsc{WeSTClass}-HAN -- New York Times]{
\includegraphics[width = 0.215\textwidth]{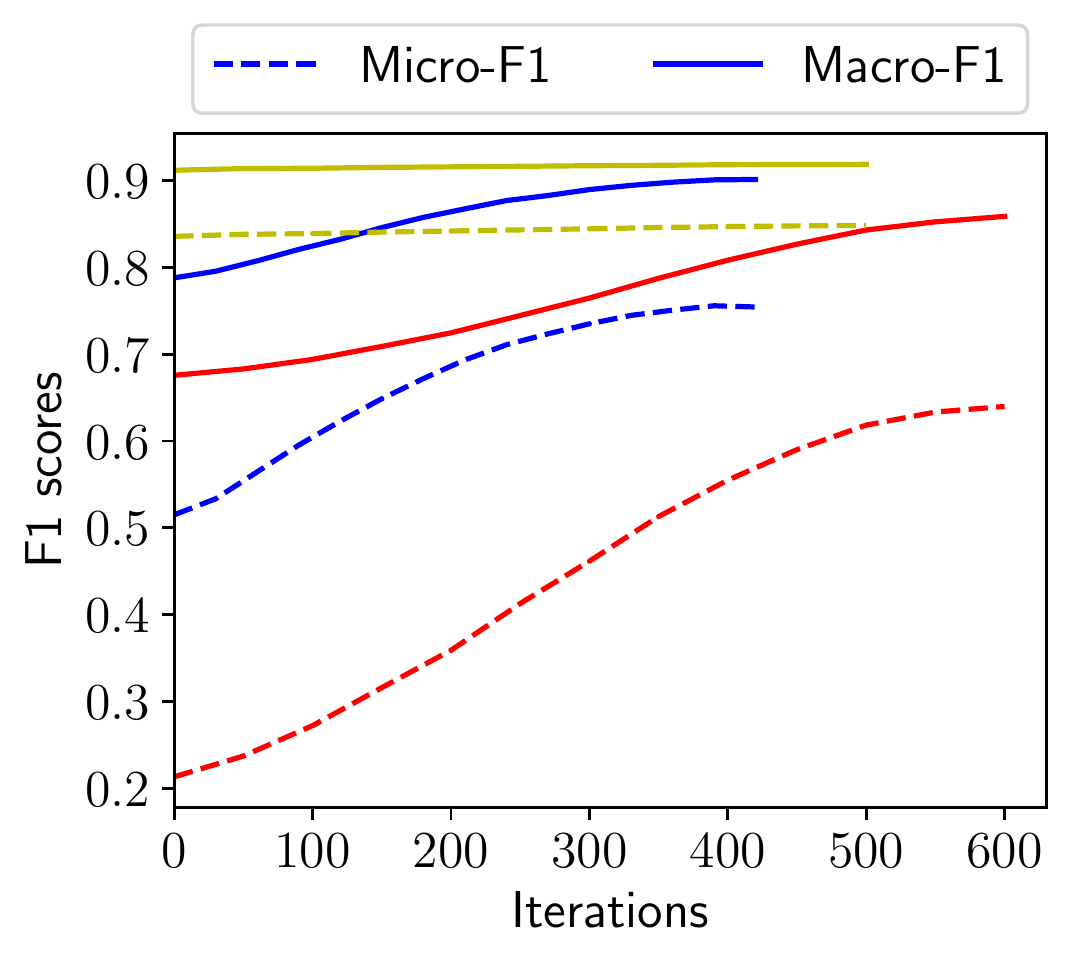}
}
\vspace{-0.1cm}
\subfigure[\textsc{WeSTClass}-CNN -- AG's News]{
\includegraphics[width = 0.225\textwidth]{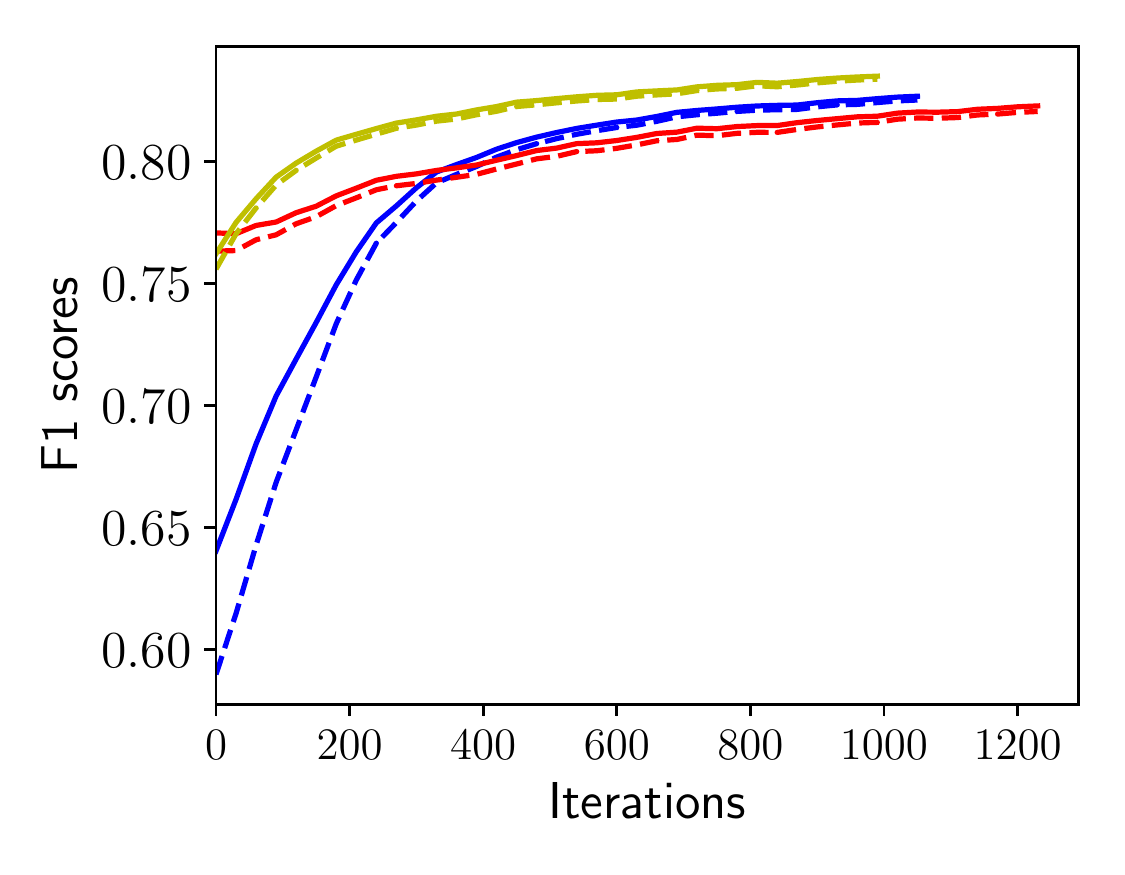}
}
\vspace{-0.1cm}
\subfigure[\textsc{WeSTClass}-HAN -- AG's News]{
\includegraphics[width = 0.225\textwidth]{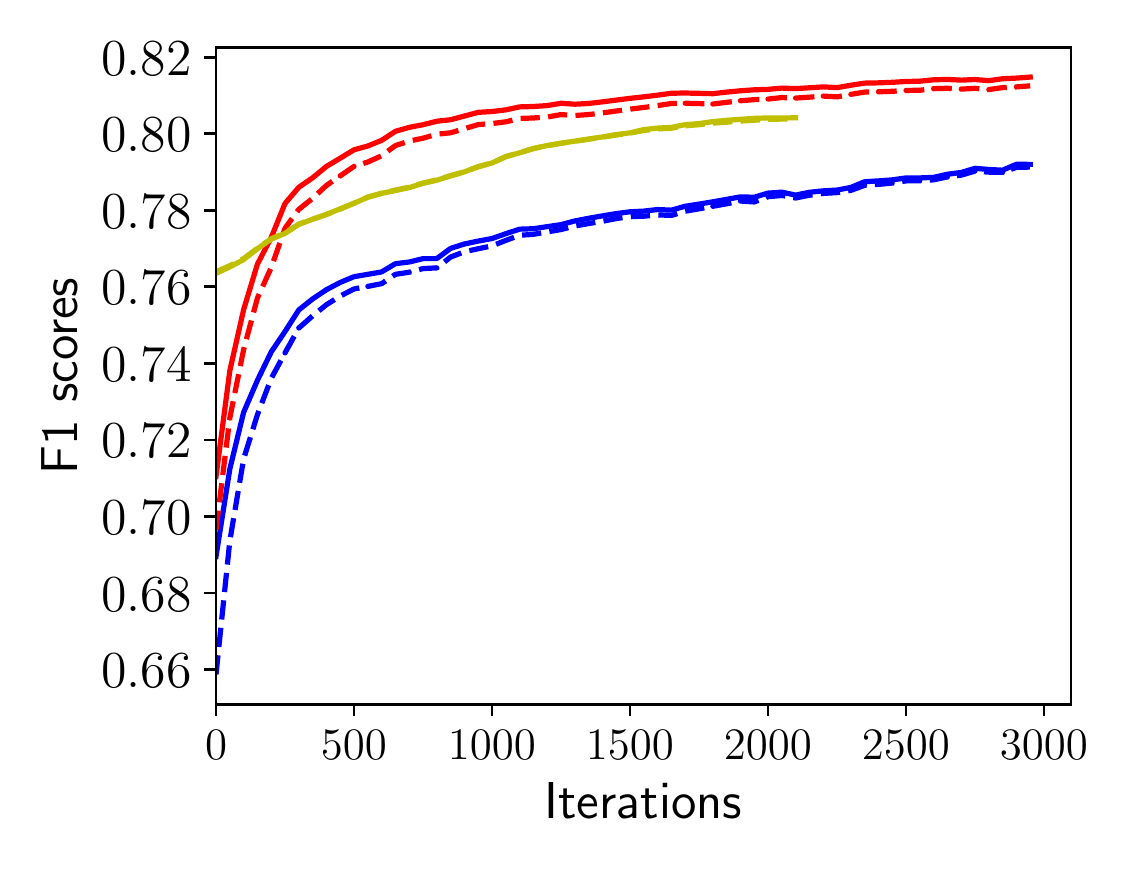}
}
\vspace{-0.1cm}
\subfigure[\textsc{WeSTClass}-CNN -- Yelp Review]{
\includegraphics[width = 0.22\textwidth]{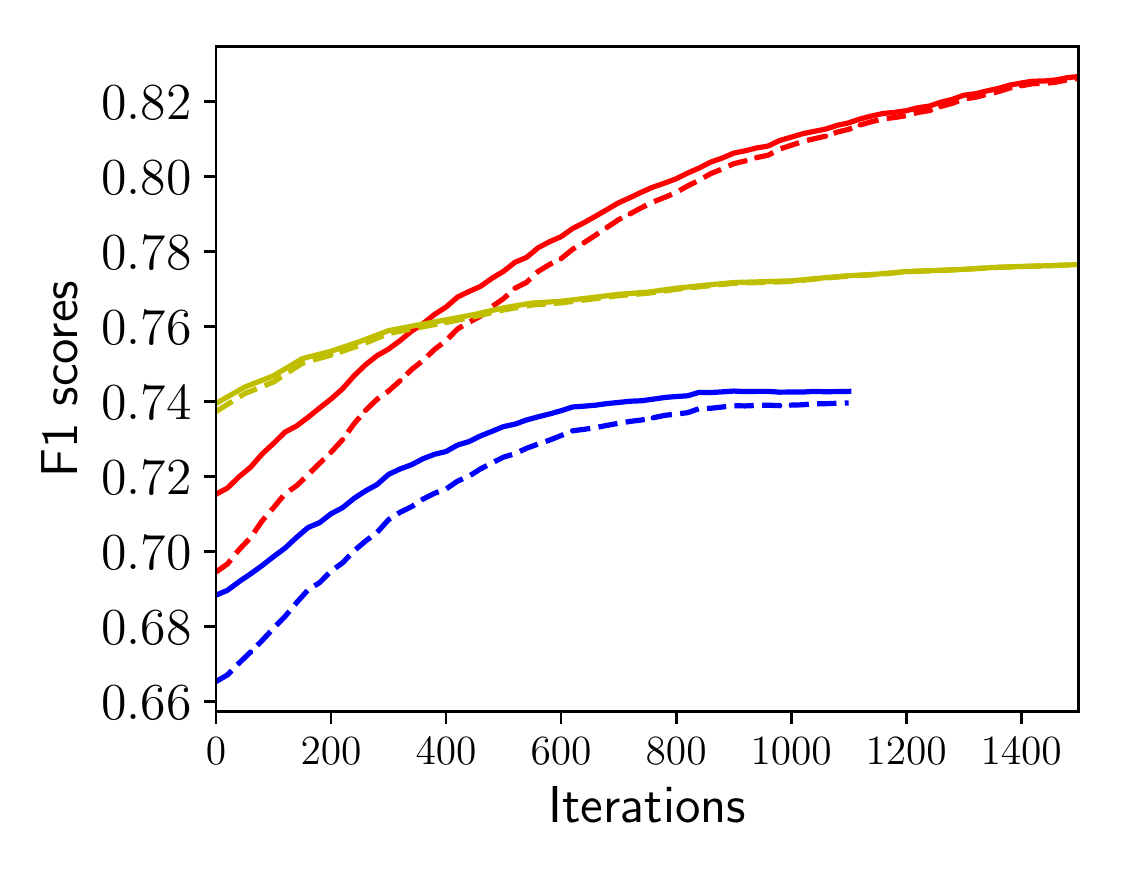}
}
\vspace{-0.1cm}
\subfigure[\textsc{WeSTClass}-HAN -- Yelp Review]{
\includegraphics[width = 0.22\textwidth]{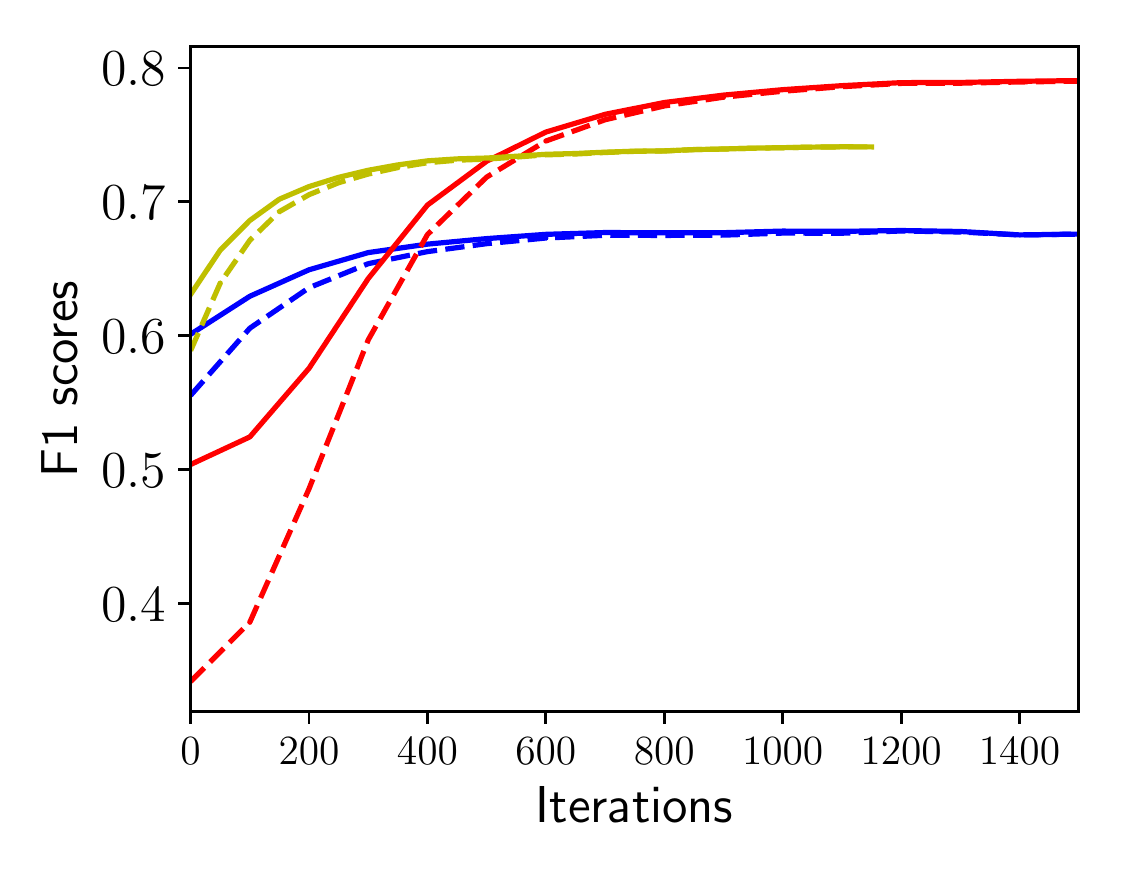}
}
\vspace{-0.3cm}
\caption{Effect of self-training modules on three datasets.}
\label{fig:self_train}
\vspace{-0.3cm}
\end{figure}

In this set of experiments, we
conduct more experiments to study the effect of self-training module in \textsc{WeSTClass},
by investigating the performance of difference models as the number of
iterations increases.  The results are shown in Figure \ref{fig:self_train}.
We can see that the self-training module can effectively improve the model
performance after the pre-training step.  Also, we find that the self-training
module generally has the least effect when supervision comes from labeled
documents.  One possible explanation is that when labeled documents are given,
we will use both pseudo documents and provided labeled documents to pre-train
the neural models. 
Such mixture training can often lead
to better model initialization, compared to using pseudo documents only.  As a
result, there is less room for self-training module to make huge improvements.

\begin{table*}
\caption{Macro-F1 scores for all methods on three datasets. \textbf{LABELS}, \textbf{KEYWORDS}, and \textbf{DOCS} means the type of seed supervision is label surface name, class-related keywords, and labeled documents, respectively.}
\vspace{-0.3cm}
\label{tab:macro}
\scalebox{0.9}{
\begin{tabular}{*{10}{c}}
\toprule
\textbf{Methods} & \multicolumn{3}{c}{\textbf{The New York Times}} & \multicolumn{3}{c}{\textbf{AG's News}} & \multicolumn{3}{c}{\textbf{Yelp Review}} \\
\cmidrule(l){2-10}
& \textbf{LABELS} & \textbf{KEYWORDS} & \textbf{DOCS} & \textbf{LABELS} & \textbf{KEYWORDS} & \textbf{DOCS} & \textbf{LABELS} & \textbf{KEYWORDS} & \textbf{DOCS}\\
\midrule
IR with tf-idf & $0.319$ & $0.509$ & - & $0.187$ & $0.258$ & - & $0.533$ & $0.638$ & -\\

Topic Model & $0.301$ & $0.253$ & - & $0.496$ & $0.723$ & - & $0.333$ & $0.333$ & - \\

Dataless & $0.484$ & - & - & $0.688$ & - & - & $0.337$ & - & -\\

UNEC & $0.690$ & - & - & $0.659$ & - & - & $0.602$ & - & -\\

PTE & - & - & $0.834\ (0.024)$ & - & - & $0.542\ (0.029)$ & - & - & $0.658\ (0.042)$\\

HAN & $0.348$ & $0.534$ & $0.740\ (0.059)$ & $0.498$ & $0.621$ & $0.731\ (0.029)$ & $0.519$ & $0.631$ & $0.686\ (0.046)$\\

CNN & $0.338$ & $0.632$ & $0.702\ (0.059)$ & $0.758$ & $0.770$ & $0.766\ (0.035)$ & $0.523$ & $0.633$ & $0.634\ (0.096)$\\

NoST-HAN & $0.515$ & $0.213$ & $0.823\ (0.035)$& $0.590$ & $0.727$ & $0.745\ (0.038)$ & $0.731$ & $0.338$ & $0.682\ (0.090)$\\

NoST-CNN & $0.701$ & $0.702$ & $0.833\ (0.013)$ & $0.534$ & $0.759$ & $0.759\ (0.032)$ & $0.639$ & $0.740$ & $0.717\ (0.058)$\\

\midrule
\textsc{WeSTClass}-HAN & {$0.754$} & {$0.640$} & $0.832\ (0.028)$ & $0.816$ & $0.820$ & $0.782\ (0.028)$ & $\mathbf{0.769}$ & $0.736$ & $0.729\ (0.040)$\\

\textsc{WeSTClass}-CNN & {$\mathbf{0.830}$} & {$\mathbf{0.837}$} & {$\mathbf{0.835}\ (\mathbf{0.010})$} & {$\mathbf{0.822}$} & {$\mathbf{0.821}$} & {$\mathbf{0.839}\ (\mathbf{0.007})$} & $0.735$ & $\mathbf{0.816}$ & $\mathbf{0.775}\ (\mathbf{0.037})$\\
\bottomrule
\end{tabular}
\vspace{0.1cm}
}
\end{table*}

\begin{table*}
\caption{Micro-F1 scores for all methods on three datasets. \textbf{LABELS}, \textbf{KEYWORDS}, and \textbf{DOCS} means the type of seed supervision is label surface name, class-related keywords, and labeled documents, respectively.}
\vspace{-0.3cm}
\label{tab:micro}
\scalebox{0.9}{
\begin{tabular}{*{10}{c}}
\toprule
\textbf{Methods} & \multicolumn{3}{c}{\textbf{The New York Times}} & \multicolumn{3}{c}{\textbf{AG's News}} & \multicolumn{3}{c}{\textbf{Yelp Review}} \\
\cmidrule(l){2-10}
& \textbf{LABELS} & \textbf{KEYWORDS} & \textbf{DOCS} & \textbf{LABELS} & \textbf{KEYWORDS} & \textbf{DOCS} & \textbf{LABELS} & \textbf{KEYWORDS} & \textbf{DOCS}\\
\midrule
IR with tf-idf & $0.240$ & $0.346$ & - & $0.292$ & $0.333$ & - & $0.548$ & $0.652$ & -\\

Topic Model & $0.666$ & $0.623$ & - & $0.584$ & $0.735$ & - & $0.500$ & $0.500$ & - \\

Dataless & $0.710$ & - & - & $0.699$ & - & - & $0.500$ & - & -\\

UNEC & $0.810$ & - & - & $0.668$ & - & - & $0.603$ & - & -\\

PTE & - & - & $0.906\ (0.020)$ & - & - & $0.544\ (0.031)$ & - & - & $0.674\ (\mathbf{0.029})$\\

HAN & $0.251$ & $0.595$ & $0.849\ (0.038)$ & $0.500$ & $0.619$ & $0.733\ (0.029)$ & $0.530$ & $0.643$ & $0.690\ (0.042)$\\

CNN & $0.246$ & $0.620$ & $0.798\ (0.085)$ & $0.759$ & $0.771$ & $0.769\ (0.034)$ & $0.534$ & $0.646$ & $0.662\ (0.062)$\\

NoST-HAN & $0.788$ & $0.676$ & $0.906\ (0.021)$& $0.619$ & $0.736$ & $0.747\ (0.037)$ & $0.740$ & $0.502$ & $0.698\ (0.066)$\\

NoST-CNN & $0.767$ & $0.780$ & $0.908\ (0.013)$ & $0.553$ & $0.766$ & $0.765\ (0.031)$ & $0.671$ & $0.750$ & $0.725\ (0.050)$\\

\midrule
\textsc{WeSTClass}-HAN & {$0.901$} & {$0.859$} & $0.908\ (0.019)$ & $0.816$ & $0.822$ & $0.782\ (0.028)$ & $\mathbf{0.771}$ & $0.737$ & $0.729\ (0.040)$\\

\textsc{WeSTClass}-CNN & {$\mathbf{0.916}$} & {$\mathbf{0.912}$} & {$\mathbf{0.911}\ (\mathbf{0.007})$} & {$\mathbf{0.823}$} & {$\mathbf{0.823}$} & {$\mathbf{0.841}\ (\mathbf{0.007})$} & $0.741$ & $\mathbf{0.816}$ & $\mathbf{0.776}\ (0.037)$\\
\bottomrule
\end{tabular}
}
\end{table*}

\subsubsection{Effect of the number of labeled documents} 
\begin{figure}
\includegraphics[width = 0.35\textwidth]{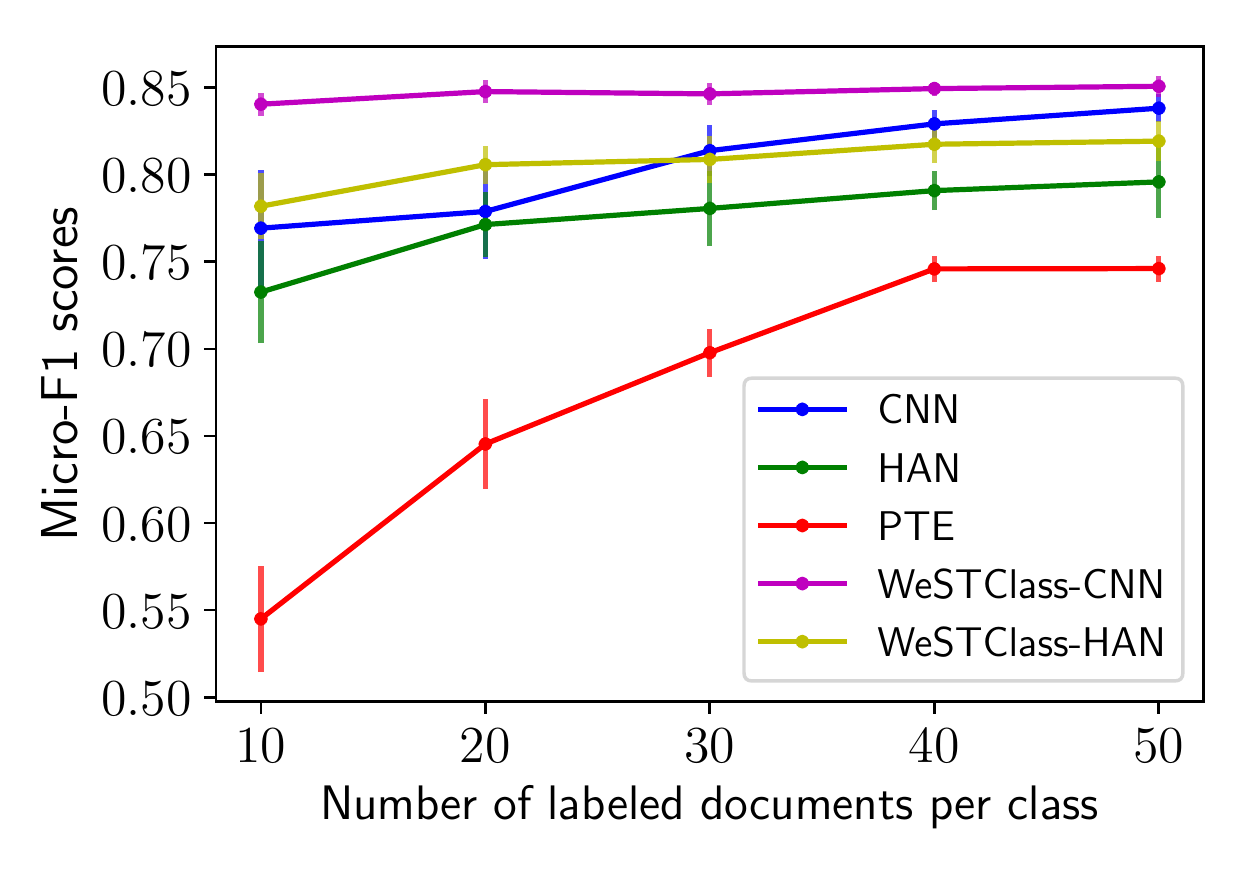}
\vspace{-0.3cm}
\caption{The performances of different methods on AG's News dataset when the
  number of labeled documents varies.}
\label{fig:num_docs}
\vspace{-0.5cm}
\end{figure}

When weak supervision
signal comes from labeled documents, the setting is similar to semi-supervised
learning except that the amount of labeled documents is very limited.  In this
set of experiments, we vary the number of labeled documents per class and
compare the performances of five methods on the AG's News dataset:
\textbf{CNN}, \textbf{HAN}, \textbf{PTE}, \textbf{\textsc{WeSTClass}-CNN} and
\textbf{\textsc{WeSTClass}-HAN}.  Again, we run each method 10 times with different sets of
labeled documents, and report the average performances with standard deviation
(represented as error bars) in Figure \ref{fig:num_docs}.  We can see that when
the amount of labeled documents is relatively large, the performances of the
five methods are comparable.  However, when fewer labeled documents are
provided, \textbf{PTE}, \textbf{CNN} and \textbf{HAN} not only exhibit obvious
performance drop, but also become very sensitive to the seed documents.
Nevertheless, \textbf{\textsc{WeSTClass}}-based models, especially \textbf{\textsc{WeSTClass}-CNN}, yield stable
performance with varying amount of labeled documents.  This phenomenon shows
that our method can more effectively take advantage of the limited amount of
seed information to achieve better performance.


\subsection{Parameter Study} In this section, we study the
effects of different hyperparameter settings on the performance of
\textbf{\textsc{WeSTClass}} with CNN and HAN models, including (1) background word
distribution weight $\alpha$, (2) number of generated pseudo documents $\beta$
for pre-training and (3) keyword vocabulary size $\gamma$ used in equation
(\ref{eq:1}) where $\gamma = |V_{d_i}|$.  When studying the effect of one
parameter, the other parameters are set to their default values as described in
Section \ref{sec:exp_setting} . We conduct all the parameter studies on the
\textbf{AG's News} dataset.

\subsubsection{Background Word Distribution Weight} 
\begin{figure}[t]
\vspace{-0.3cm}
\subfigure[\textsc{WeSTClass}-CNN]{
\includegraphics[width = 0.22\textwidth]{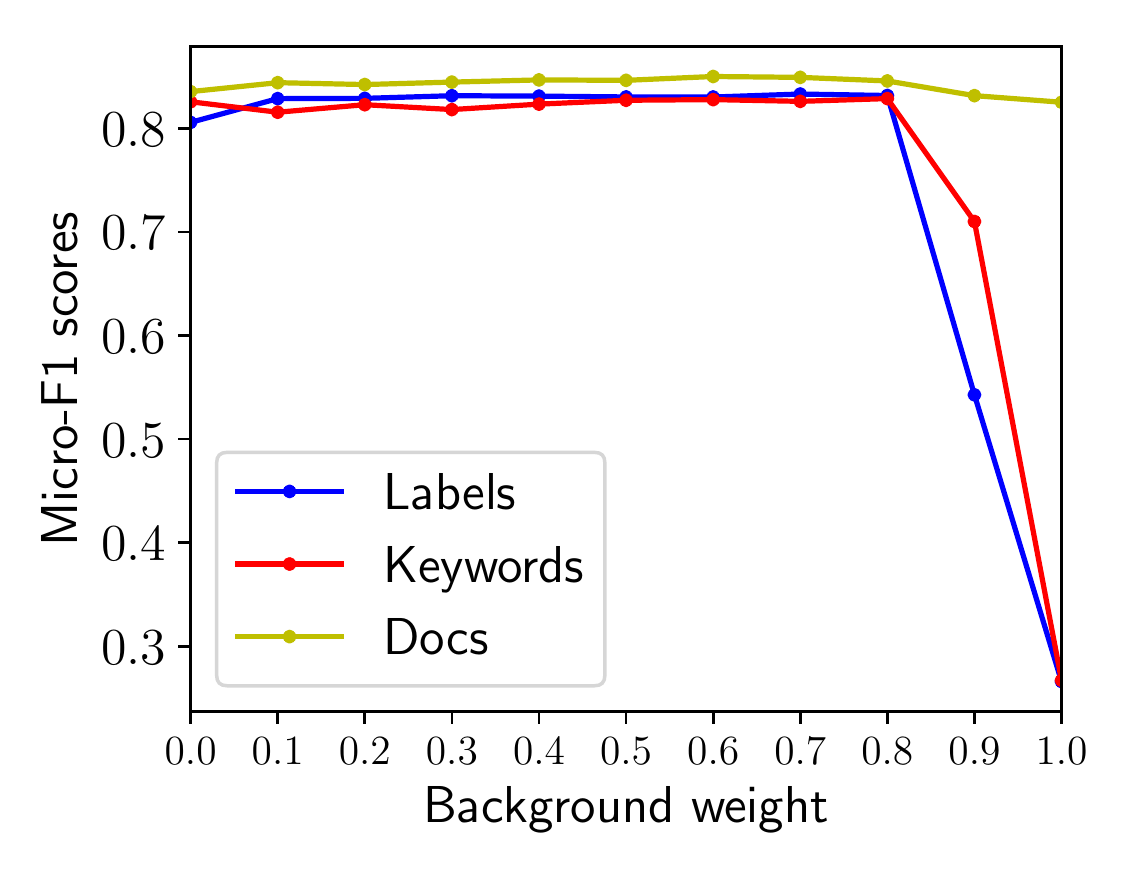}
\label{fig:background_cnn}
}
\subfigure[\textsc{WeSTClass}-HAN]{
\includegraphics[width = 0.22\textwidth]{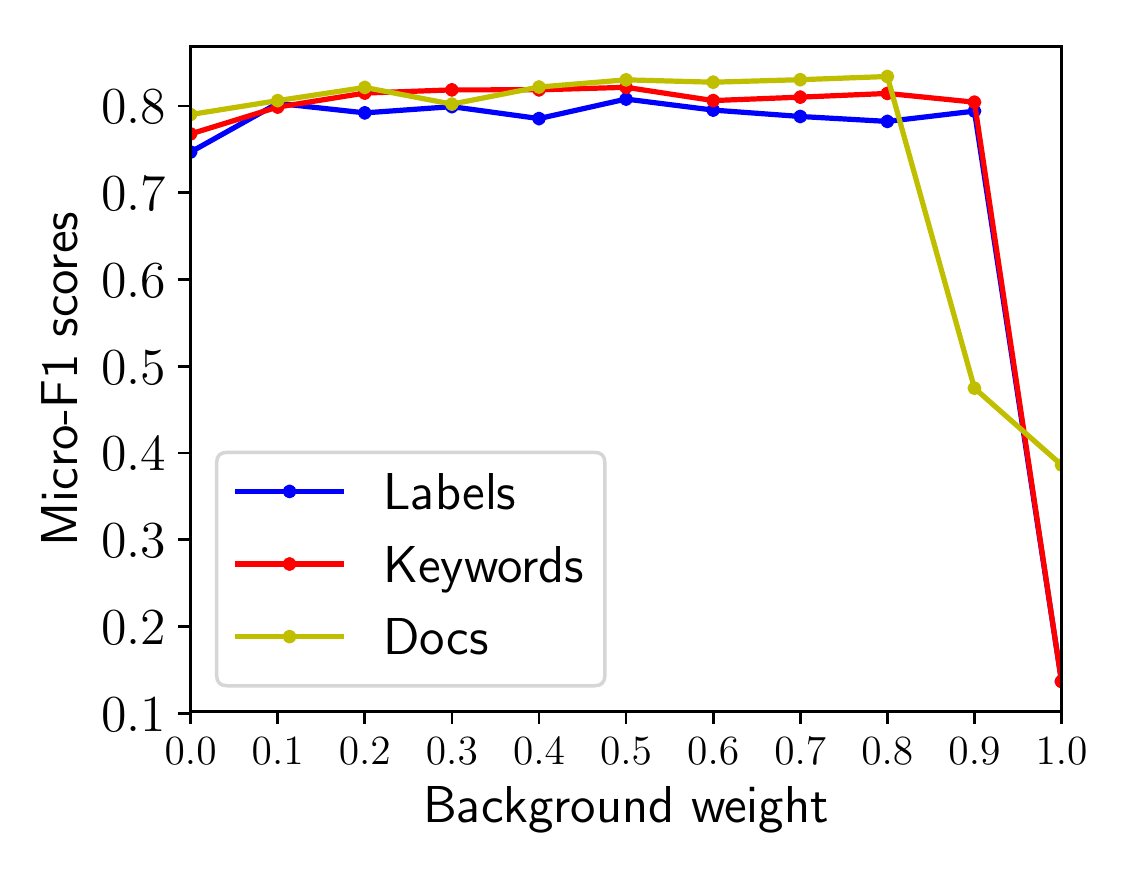}
}
\vspace{-0.5cm}
\caption{Effect of background word distribution weight $\alpha$ on AG's News dataset.}
\label{fig:background}
\end{figure}

The background word
distribution weight $\alpha$ is used in both the language model for pseudo
documents generation and pseudo-labels computation.  When $\alpha$ becomes
smaller, the generated pseudo documents contain more topic-related words and
fewer background words, and the pseudo-labels become similar to one-hot
encodings.  We vary $\alpha$ from $0$ to $1$ with interval equal to $0.1$.  The
effect of $\alpha$ is shown in Figure \ref{fig:background}.  Overall, different
$\alpha$ values result in comparable performance, except when $\alpha$ is close
to $1$, pseudo documents and pseudo-labels become uninformative: pseudo
documents are generated directly from background word distribution without any
topic-related information, and pseudo-labels are uniform distributions.  We
notice that when $\alpha=1$, \textbf{labeled documents} as supervision source
results in much better performance than \textbf{label surface name} and
\textbf{class-related keywords}.  This is because pre-training with
\textbf{labeled documents} is performed using both pseudo documents and labeled
documents, and the provided labeled documents are still informative.  When
$\alpha$ is close to $0$, the performance is slightly worse than other
settings, because pseudo documents only contain topic-related keywords and
pseudo-labels are one-hot encodings, which can easily lead to model overfitting
to pseudo documents and behaving worse on real documents classification.

\subsubsection{Number of pseudo documents for pre-training} 
\begin{figure}[t]
\vspace{-0.38cm}
\subfigure[\textsc{WeSTClass}-CNN]{
\includegraphics[width = 0.22\textwidth]{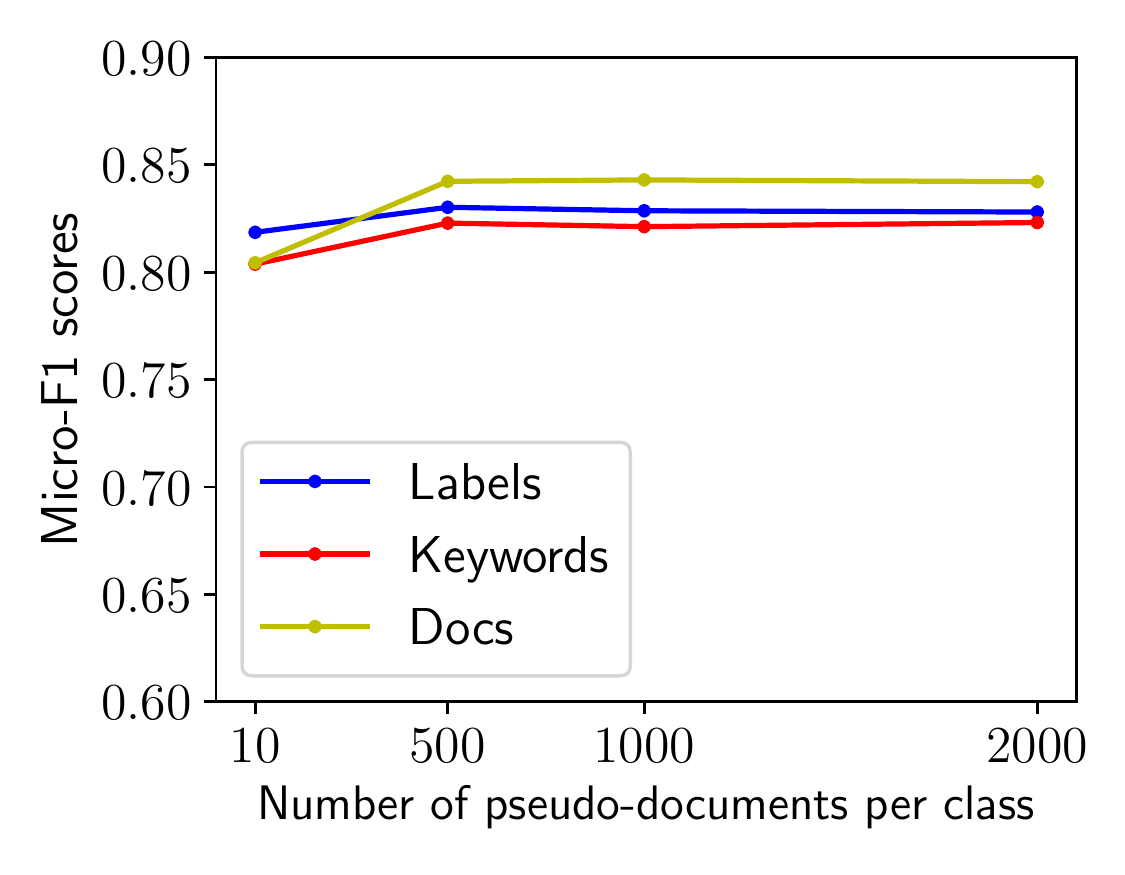}
}
\subfigure[\textsc{WeSTClass}-HAN]{
\includegraphics[width = 0.22\textwidth]{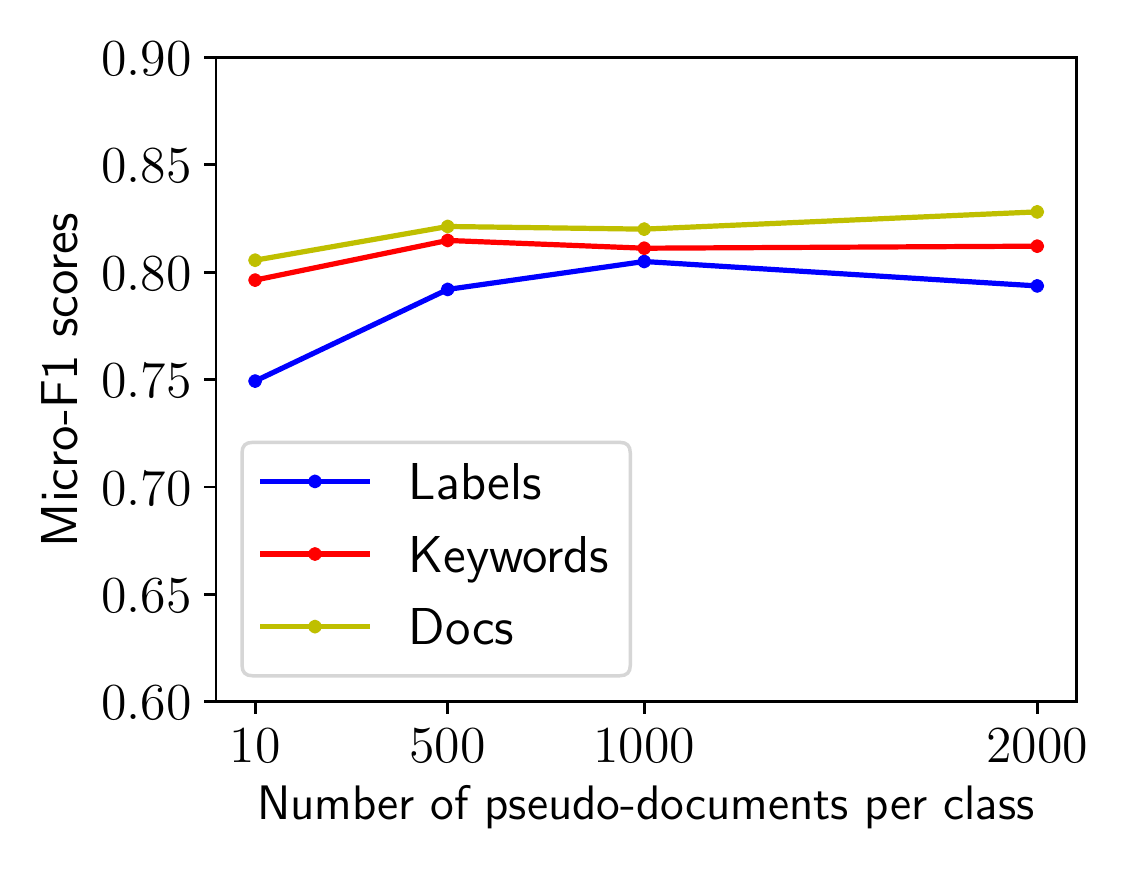}
}
\vspace{-0.3cm}
\caption{Effect of pseudo documents amount per class $\beta$ for pre-training on AG's News dataset.}
\label{fig:num_doc}
\vspace{-0.3cm}
\end{figure}

The effect of
pseudo documents amount $\beta$ is shown in Figure \ref{fig:num_doc}. We have
the following findings from Figure \ref{fig:num_doc}:  On the one hand, if the
amount of generated pseudo documents is too small, the information carried in
pseudo documents will be insufficient to pre-train a good model. On the other
hand, generating too many pseudo documents will make the pre-training process
unnecessarily long. Generating $500$ to $1000$ pseudo documents of each class
for pre-training will strike a good balance between pre-training time and model
performance.

\subsubsection{Size of Keyword Vocabulary} 
\begin{figure}[t]
\vspace{-0.3cm}
\subfigure[\textsc{WeSTClass}-CNN]{
\includegraphics[width = 0.22\textwidth]{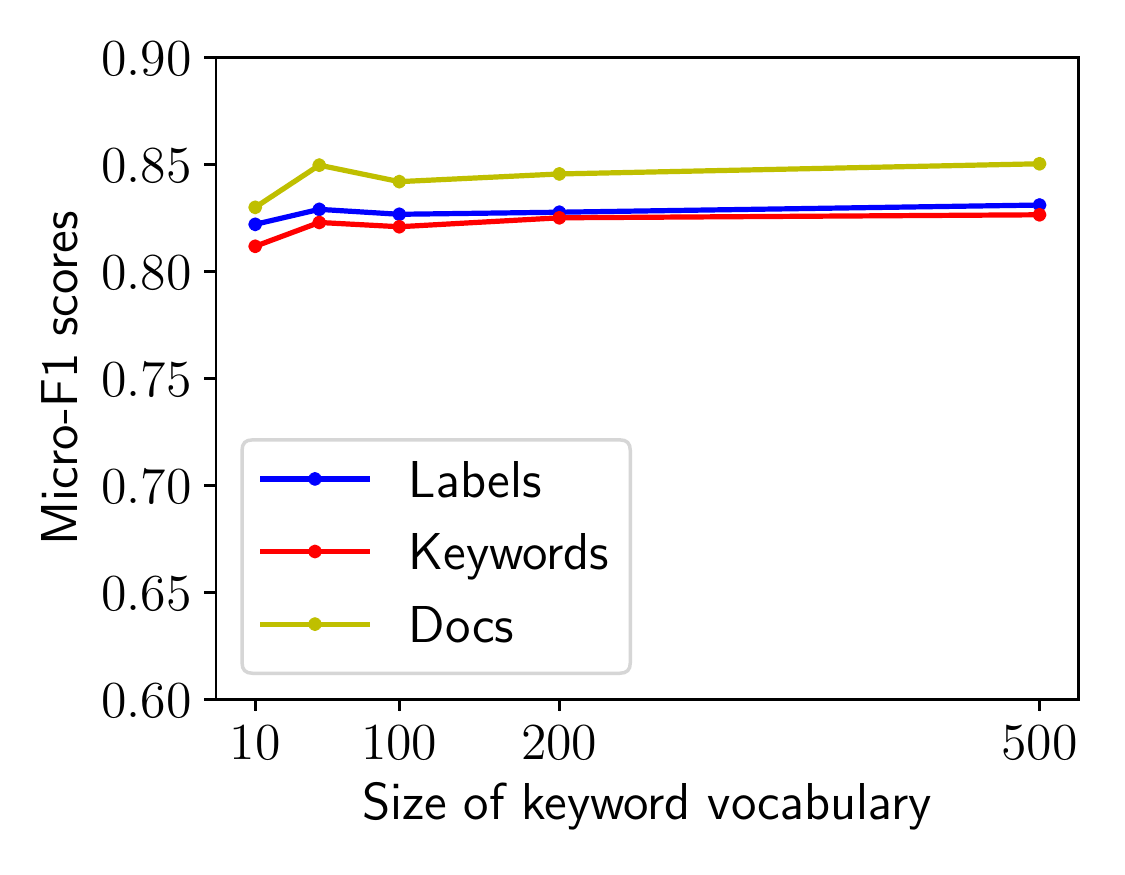}
}
\subfigure[\textsc{WeSTClass}-HAN]{
\includegraphics[width = 0.22\textwidth]{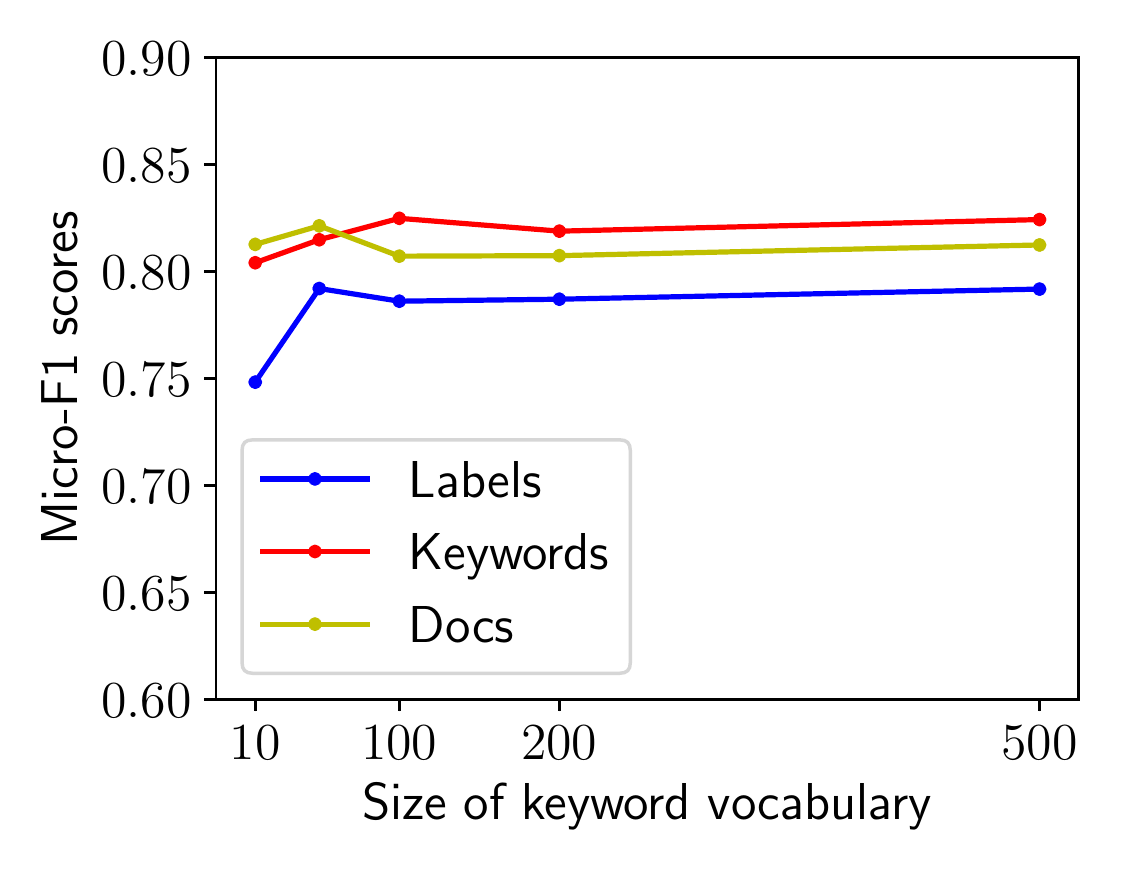}
}
\vspace{-0.5cm}
\caption{Effect of keyword vocabulary size $\gamma$ on AG's News dataset.}
\label{fig:total_num}
\end{figure}

Recall the pseudo document
generation process in Section \ref{sec:gen}, after sampling a document vector
$\bs{d}_{i}$, we will first construct a keyword vocabulary $V_{d_{i}}$ that
contains the top-$\gamma$ words with most similar word embedding with
$\bs{d}_i$.  The size of the keyword vocabulary $\gamma$ controls the number of
unique words that appear frequently in the generated pseudo documents. If
$\gamma$ is too small, only a few topical keywords will appear frequently in
pseudo documents, which will reduce the generalization ability of the
pre-trained model. As shown in Figure \ref{fig:total_num}, $\gamma$ can be
safely set within a relatively wide range from $50$ to $500$ in practice.

\subsection{Case Study}

In this subsection, we perform a set of case studies to further understand the
properties of our proposed method.

\begin{table*}[!t]
\caption{Keyword Lists at Top Percentages of Average Tf-idf.}
\vspace{-0.3cm}
\label{tab:keywords}
\scalebox{0.8}{
\begin{tabular}{*{4}{c}}
\toprule
Class & $1 \%$ & $5 \%$ & $10 \%$ \\
\midrule
Politics & \{government, president, minister\} & \{mediators, criminals, socialist\} & \{suspending, minor, lawsuits\}\\
Sports & \{game, season, team\} & \{judges, folks, champagne\} & \{challenging, youngsters, stretches\}\\
Business & \{profit, company, sales\} & \{refunds, organizations, trader\} & \{winemaker, skilling, manufactured\}\\
Technology & \{internet, web, microsoft\} & \{biologists, virtually, programme\} & \{demos, microscopic, journals\}\\
\bottomrule
\end{tabular}
}
\end{table*}

\begin{figure}[t]
\vspace{-0.5cm}
\subfigure[]{
	\label{fig:keywords}
	\includegraphics[width = 0.216\textwidth]{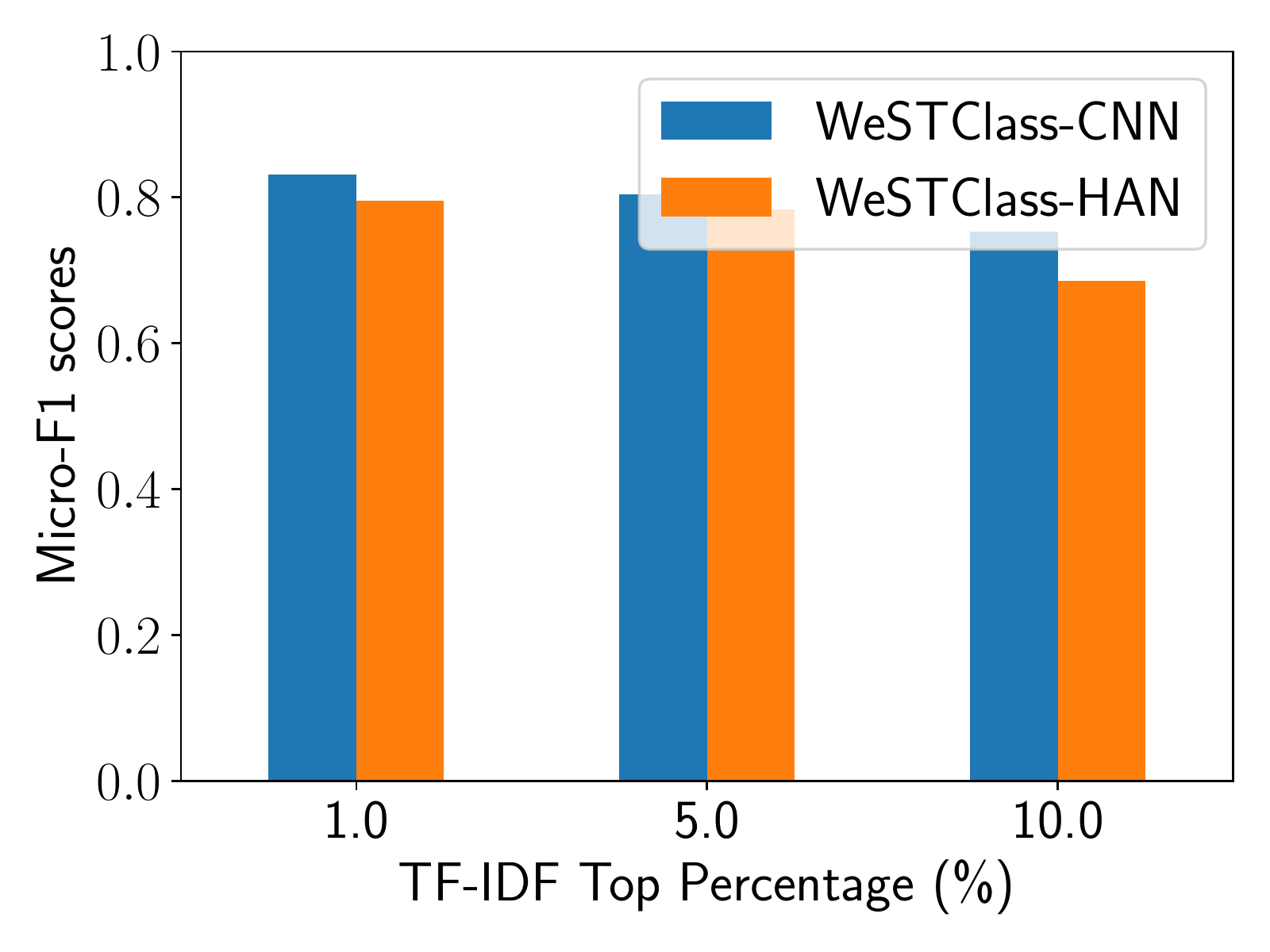}
}
\subfigure[]{
	\label{fig:case_study}
	\includegraphics[width = 0.234\textwidth]{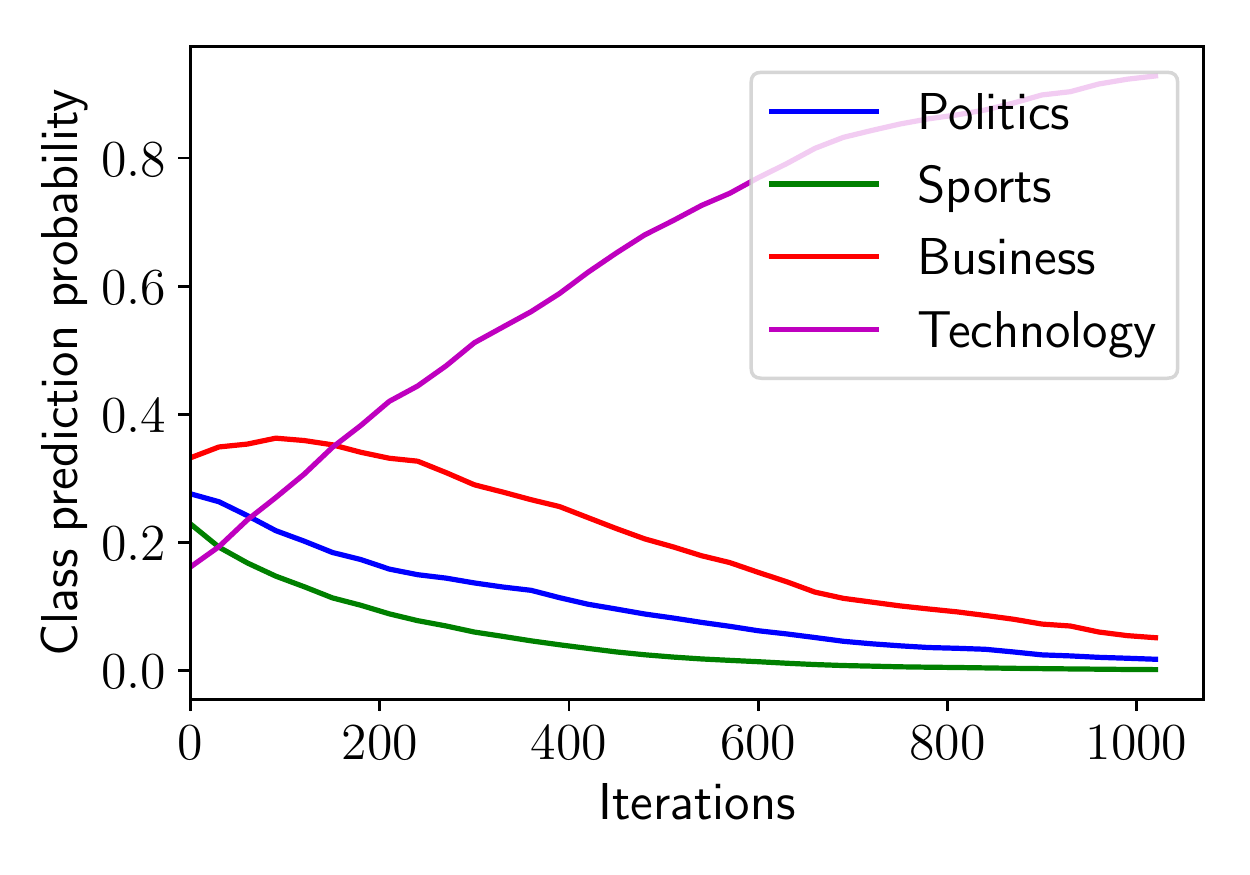}
}
\vspace{-0.3cm}
\caption{(a) Performances on AG's News dataset under different sets of seed keywords. (b) Class prediction probability during self-training procedure for a sample document. }
\label{fig:cases}
\vspace{-0.4cm}
\end{figure}



\subsubsection{Choice of Seed Keywords}
\label{subsec:keyword_sensitivity}
In the first set of case studies, we are interested in how sensitive our model
is to the selection of seed keywords.  In Section \ref{sec:exp_setting}, we
manually select class-related keywords, which could be subjective.  Here we
explore the sensitivity of \textbf{\textsc{WeSTClass}-CNN} and \textbf{\textsc{WeSTClass}-HAN} to different
sets of seed keywords.  For each class $j$ of \textbf{AG's News} dataset, we
first collect all documents
belonging to class $j$, and then compute the tf-idf weighting of each word in each document of class $j$. We sort each word's average tf-idf weighting in these documents from high to low. Finally we form the seed keyword
lists by finding words that rank at top $1\%$ (most relevant), $5\%$ and $10\%$ based on the average tf-idf value. The keywords of each class at these percentages are shown in Table
\ref{tab:keywords}; the performances of \textbf{\textsc{WeSTClass}-CNN} and \textbf{\textsc{WeSTClass}-HAN}
are shown in Figure \ref{fig:keywords}. At top $5\%$ and $10\%$ of the average tf-idf weighting,
although some keywords are already slightly irrelevant to their corresponding
class semantic, \textbf{\textsc{WeSTClass}-CNN} and \textbf{\textsc{WeSTClass}-HAN} still perform
reasonably well, which shows the robustness of our proposed framework to
different sets of seed keywords.

\subsubsection{Self-training Corrects Misclassification}
In the second set of case studies, we are interested in how the self-training
module behaves to improve the performance of our model.  Figure
\ref{fig:case_study} shows \textbf{\textsc{WeSTClass}-CNN}'s prediction with \textbf{label
  surface name} as supervision source on a sample document from \textbf{AG's
  News} dataset: \emph{The national competition regulator has elected not to
  oppose Telstra's 3G radio access network sharing arrangement with rival telco
  Hutchison.} We notice that this document is initially misclassified after the
pre-training procedure, but it is then corrected by the subsequent
self-training step.  This example shows that neural models have the ability of
self-correcting by learning from its high-confidence predictions with
appropriate pre-training initialization.



\section{Discussions and Conclusions}

We have proposed a weakly-supervised text classification method
built upon neural classifiers. With (1) a pseudo document
generator for generating pseudo training data and (2) a
self-training module that bootstraps on real unlabled data for
model refining, our method effectively addresses the key
bottleneck for existing neural text classifiers---the lack of
labeled training data. Our method is not only flexible in
incorporating difference sources of weak supervision (class
label surface names, class-related keywords, and labeled
documents), but also generic enough to support different neural
models (CNN and RNN).  Our experimental results have shown that
our method outperforms baseline methods significantly, and it
is quite robust to different settings of hyperparameters and
different types of user-provided seed information.

An interesting finding based on the experiments in Section \ref{sec:exp} is
that different types of weak supervision are all highly helpful for 
the good performances of neural models. In the future, it is interesting to study how
to effectively integrate different types of seed information to further boost the
performance of our method.




\section*{Acknowledgements}
This research is sponsored in part by U.S. Army Research Lab. under Cooperative Agreement No. W911NF-09-2-0053 (NSCTA), DARPA under Agreement No. W911NF-17-C-0099, National Science Foundation IIS 16-18481, IIS 17-04532, and IIS-17-41317, DTRA HDTRA11810026, and grant 1U54GM114838 awarded by NIGMS through funds provided by the trans-NIH Big Data to Knowledge (BD2K) initiative (www.bd2k.nih.gov). We thank anonymous reviewers for valuable and insightful feedback.

\bibliographystyle{ACM-Reference-Format}
\bibliography{ref}

\end{document}